\documentclass[authorversion,sigconf,bookmarksnumbered,unicode,bookmarks=false,hidelinks]{acmart}

\usepackage{tabularx}
\usepackage{ragged2e}
\usepackage[font=small]{caption} 
\newcolumntype{L}[1]{>{\hsize=#1\hsize\RaggedRight} X}
\usepackage[frozencache=true,cachedir=minted-cache]{minted}
\AtBeginEnvironment{minted}{%
  }
\AtBeginDocument{%
  \providecommand\BibTeX{{%
    \normalfont B\kern-0.5em{\scshape i\kern-0.25em b}\kern-0.8em\TeX}}}
\usepackage{caption}
\usepackage{subcaption}
\usepackage[inline]{enumitem}

\usepackage{color, colortbl}
\definecolor{Gray}{gray}{0.9}
\usepackage{soul}

\copyrightyear{2022}
\acmYear{2022}
\setcopyright{acmlicensed}\acmConference[ASE '22]{37th IEEE/ACM
International Conference on Automated Software Engineering}{October 10--14,
2022}{Rochester, MI, USA}
\acmBooktitle{37th IEEE/ACM International Conference on Automated Software
Engineering (ASE '22), October 10--14, 2022, Rochester, MI, USA}
\acmPrice{15.00}
\acmDOI{10.1145/3551349.3556945}
\acmISBN{978-1-4503-9475-8/22/10}

\begin{document}

%%
%% The "title" command has an optional parameter,
%% allowing the author to define a "short title" to be used in page headers.
\def\GLITCH{GLITCH}
%\title{\GLITCH: an Intermediate-Representation-Based Security Analysis for Infrastructure as Code Scripts}
%\title{An empirical study on security smells across multiple IaC technologies using the polyglot tool \GLITCH}
%\title{Automated Polyglot Security Smell Detection in Infrastructure as Code: A Large Empirical Investigation}
%\title{Automated Polyglot Security Smell Detection in Infrastructure as Code}
%\title{Security Smells across Multiple Infrastructure as Code Technologies: a Polyglot Approach to Smell Detection}
\title{\GLITCH: Automated Polyglot Security Smell Detection in Infrastructure as Code}

%%
%% The "author" command and its associated commands are used to define
%% the authors and their affiliations.
%% Of note is the shared affiliation of the first two authors, and the
%% "authornote" and "authornotemark" commands
%% used to denote shared contribution to the research.
\author{Nuno Saavedra}
\email{nuno.saavedra@tecnico.ulisboa.pt}
%\orcid{1234-5678-9012}
\affiliation{%
  \institution{INESC-ID and IST, University of Lisbon}
%  \streetaddress{P.O. Box 1212}
  \city{Lisbon}
%  \state{Ohio}
  \country{Portugal}
%  \postcode{43017-6221}
}

\author{Jo\~{a}o F. Ferreira}
\email{joao@joaoff.com}
%\orcid{1234-5678-9012}
\affiliation{%
  \institution{INESC-ID and IST, University of Lisbon}
%  \streetaddress{P.O. Box 1212}
  \city{Lisbon}
%  \state{Ohio}
  \country{Portugal}
%  \postcode{43017-6221}
}

%\author{Ben Trovato}
%\authornote{Both authors contributed equally to this research.}
%\email{trovato@corporation.com}
%\orcid{1234-5678-9012}
%\author{G.K.M. Tobin}
%\authornotemark[1]
%\email{webmaster@marysville-ohio.com}
%\affiliation{%
%  \institution{Institute for Clarity in Documentation}
%  \streetaddress{P.O. Box 1212}
%  \city{Dublin}
%  \state{Ohio}
%  \country{USA}
%  \postcode{43017-6221}
%}

%\author{Lars Th{\o}rv{\"a}ld}
%\affiliation{%
%  \institution{The Th{\o}rv{\"a}ld Group}
%  \streetaddress{1 Th{\o}rv{\"a}ld Circle}
%  \city{Hekla}
%  \country{Iceland}}
%\email{larst@affiliation.org}

%\author{Valerie B\'eranger}
%\affiliation{%
%  \institution{Inria Paris-Rocquencourt}
%  \city{Rocquencourt}
%  \country{France}
%}

%%
%% By default, the full list of authors will be used in the page
%% headers. Often, this list is too long, and will overlap
%% other information printed in the page headers. This command allows
%% the author to define a more concise list
%% of authors' names for this purpose.
\renewcommand{\shortauthors}{Saavedra and Ferreira}

%%
%% The abstract is a short summary of the work to be presented in the
%% article.

\begin{abstract}
Infrastructure as Code (IaC) is the process of managing IT infrastructure via programmable configuration files (also called IaC scripts). Like other software artifacts, IaC scripts may contain security smells, which are coding patterns that can result in security weaknesses. Automated analysis tools to detect security smells in IaC scripts exist, but they focus on specific technologies such as Puppet, Ansible, or Chef. This means that when the detection of a new smell is implemented in one of the tools, it is not immediately available for the technologies supported by the other tools\,---\,the only option is to duplicate the effort.

This paper presents an approach that enables consistent security smell detection across different IaC technologies. We conduct a large-scale empirical study that analyzes security smells on three large datasets containing 196,755 IaC scripts and 12,281,251 LOC. We show that all categories of security smells are identified across all datasets and we identify some smells that might affect many IaC projects.
 To conduct this study, we developed \GLITCH, a new technology-agnostic framework that enables automated  polyglot smell detection by transforming IaC scripts into an intermediate representation, on which different security smell detectors can be defined. \GLITCH\ currently supports the detection of nine different security smells in scripts written in Ansible, Chef, or Puppet. We compare \GLITCH\ with state-of-the-art security smell detectors. The results obtained not only show that \GLITCH\ can reduce the effort of writing security smell analyses for multiple IaC technologies, but also that it has higher precision and recall than the current state-of-the-art tools.

%This paper presents \GLITCH, a new technology-agnostic framework that enables automated polyglot smell detection by transforming IaC scripts into an intermediate representation, on which different security smell detectors can be defined. \GLITCH\ currently supports the detection of nine different security smells in scripts written in Puppet, Ansible, or Chef. We compare \GLITCH\ with state-of-the-art security smell detectors. The results obtained not only show that \GLITCH\ can reduce the effort of writing security smell analyses for multiple IaC technologies, but also that it has higher precision and recall than the current state-of-the-art tools.

%Infrastructure as Code (IaC) is the process of managing IT infrastructure via programmable configuration files. IaC has progressively gained more adoption in the DevOps landscape. Even so, IaC is not a silver bullet; akin to other software artifacts, IaC scripts can suffer from bugs. The scientific community has been active in proposing analysis methods which can mitigate the lifespan of such errors. However, the IaC technologies ecosystem is scattered and developing a tool for one technology will not solve the industry's problems for another. In this article, we describe a technology-agnostic framework, called GLITCH, on top of which static analysis techniques can be developed. The framework uses an intermediate representation to which scripts from multiple technologies can be translated to. We used our framework to implement an analysis technique to detect security smells in IaC scripts based on previous work. TBW - Results
\end{abstract}

%%
%% The code below is generated by the tool at http://dl.acm.org/ccs.cfm.
%% Please copy and paste the code instead of the example below.
%%

%\begin{CCSXML}
%<ccs2012>
%   <concept>
%       <concept_id>10002978.10003022.10003023</concept_id>
%       <concept_desc>Security and privacy~Software security engineering</concept_desc>
%       <concept_significance>500</concept_significance>
%       </concept>
% </ccs2012>
%\end{CCSXML}

%\ccsdesc[500]{Security and privacy~Software security engineering}

\begin{CCSXML}
<ccs2012>
   <concept>
       <concept_id>10011007.10011006.10011071</concept_id>
       <concept_desc>Software and its engineering~Software configuration management and version control systems</concept_desc>
       <concept_significance>500</concept_significance>
       </concept>
   <concept>
       <concept_id>10011007.10011006.10011073</concept_id>
       <concept_desc>Software and its engineering~Software maintenance tools</concept_desc>
       <concept_significance>500</concept_significance>
       </concept>
   <concept>
       <concept_id>10002978.10003022</concept_id>
       <concept_desc>Security and privacy~Software and application security</concept_desc>
       <concept_significance>500</concept_significance>
       </concept>
 </ccs2012>
\end{CCSXML}

\ccsdesc[500]{Software and its engineering~Software configuration management and version control systems}
\ccsdesc[500]{Software and its engineering~Software maintenance tools}
\ccsdesc[500]{Security and privacy~Software and application security}

%%\begin{CCSXML}
%%<ccs2012>
%%<concept>
%%<concept_id>10011007.10011006.10011066.10011067</concept_id>
%%<concept_desc>Software and its engineering~Object oriented frameworks</concept_desc>
%%<concept_significance>100</concept_significance>
%%</concept>
%%<concept>
%%<concept_id>10011007.10011006.10011050.10011017</concept_id>
%%<concept_desc>Software and its engineering~Domain specific languages</concept_desc>
%%<concept_significance>500</concept_significance>
%%</concept>
%%<concept>
%%<concept_id>10011007.10011074.10011099.10011102</concept_id>
%%<concept_desc>Software and its engineering~Software defect analysis</concept_desc>
%%<concept_significance>300</concept_significance>
%%</concept>
%%<concept>
%%<concept_id>10011007.10010940.10010941.10010942</concept_id>
%%<concept_desc>Software and its engineering~Software infrastructure</concept_desc>
%%<concept_significance>500</concept_significance>
%%</concept>
%%</ccs2012>
%%\end{CCSXML}

%%\ccsdesc[500]{Software and its engineering~Software infrastructure}
%%\ccsdesc[500]{Software and its engineering~Domain specific languages}
%%\ccsdesc[300]{Software and its engineering~Software defect analysis}
%%\ccsdesc[100]{Software and its engineering~Object oriented frameworks}
%%
%% Keywords. The author(s) should pick words that accurately describe
%% the work being presented. Separate the keywords with commas.
\keywords{devops, infrastructure as code, security smells, Ansible, Chef, Puppet, intermediate model, static analysis}

%%
%% This command processes the author and affiliation and title
%% information and builds the first part of the formatted document.
\maketitle

%%
%% Introduction
\section{Introduction}
\label{sec:introduction}
Infrastructure as Code (IaC) is a process which has been progressively gaining more adoption in the DevOps world since it facilitates the provision of scalable and reproducible environments. In current practice, the use of IaC scripts is essential to efficiently maintain servers and development environments. For example, according to Rahman et al.~\cite{rahman2019seven}, Intercontinental Exchange (ICE), a Fortune 500 company, maintains 75\% of its 20,000 servers using IaC scripts. The use of IaC scripts has helped ICE decrease the time needed to provision development environments from 1--2 days to 21 minutes.

Despite its many benefits, IaC scripts may contain defects that can have serious implications. 
For instance, due to bugs in their IaC scripts, GitHub experienced an outage of their DNS infrastructure~\cite{fryman_2014} and Amazon Web Services lost around 150 million USD after issues with their S3 billing system~\cite{hersher_2017}.
To address this, there has been an effort by the research community to categorize and identify defects, and in particular, so-called security smells, which are coding patterns that can result in security weaknesses~\cite{rahman2018characterizing,rahman2019seven,rahman2020gang,rahman2021security}.
%For example, Figure~\ref{fig:iac-smell-1} presents 
% Ainda nao sabemos se mostramos este exemplo, ou apenas os outros.

Even when a security smell does not lead to a security breach, it deserves attention and inspection. It is thus important to develop automated methods that can assist developers identifying security smells in their IaC scripts. Two influential automated tools developed by the research community are SLIC~\cite{rahman2019seven}, which supports seven types of security smells in Puppet\footnote{\url{https://puppet.com}} scripts, and SLAC~\cite{rahman2021security}, which supports nine types of security smells in Chef\footnote{\url{https://www.chef.io}} scripts and six types in Ansible\footnote{\url{https://www.ansible.com}} scripts. These tools are very valuable, since they cover a wide range of security smells and three of the major IaC technologies. However, their implementations are separate and involve substantial duplication. If one wishes to implement the detection of a new smell, one has to develop a different implementation for each of the IaC technologies supported. 
Consequently, it is often the case that the detection of security smells is inconsistent for different IaC technologies. Figure~\ref{fig:chef-no-smell-ex1} presents part of a Chef script with no security smells taken from the project Vagrant Chef for CakePHP.\footnote{\url{https://github.com/FriendsOfCake/vagrant-chef/blob/288336e506a5009ed93c06a784fa93e30a27040c/cookbooks/percona/recipes/server.rb\#L28}}
For this example, SLAC reports a false positive: a non-existent security smell of type \emph{Hard-coded secret}. On the other hand, if we consider the same script in Puppet (Figure~\ref{fig:puppet-no-smell-ex1}), SLIC will not report any security smell. Surprisingly, inconsistencies exist even when considering the same tool: SLAC will not report any security smell when considering the same script in Ansible (this happens because SLAC uses separate code for Ansible and Chef).

These inconsistencies would not occur if we had 
%holistic and 
polyglot defect prediction and debugging environments for IaC, a direction recently proposed by Alnafessah et al.~\cite{alnafessah2021quality}. 
Also, a problem that has been observed by Guerriero et al., after interviews with IaC experts, is that the 
IaC technology ecosystem is currently very scattered and heterogeneous~\cite{guerriero2019adoption}. 
%and not fully understood~\cite{guerriero2019adoption}. 
Guerriero et al. also identified the need to develop more 
IaC development tools, such as static analysis tools and security-related tools.
It is thus clear that it would be beneficial to develop unifying methods that can reduce inconsistencies.
%It would thus be beneficial to develop automated methods that can be applied to different IaC tecnologies.

\begin{figure*}
     \centering
     \begin{subfigure}[b]{0.47\textwidth}
         \centering
         \footnotesize
\begin{minted}[]{ruby}
server_root_password = node['mysql']['server_root_password']
execute 'set-mysql-root' do
  command <<-EOH
    mysqladmin -u root password #{server_root_password}
    mysql -uroot -p#{server_root_password} -e (...)
  EOH
  only_if "/usr/bin/mysql -u root -e 'show databases;'"
end
\end{minted}
         \caption{Part of a Chef script (from Vagrant Chef for CakePHP)}
         \label{fig:chef-no-smell-ex1}
     \end{subfigure}
     ~~~ %\hfill
     \begin{subfigure}[b]{0.47\textwidth}
         \centering
         \footnotesize
\begin{minted}[]{ruby}
$server_root_password = $facts['mysql']['server_root_password']
exec { 'set-mysql-root':
  command => @("COMMAND"/L)
    mysqladmin -u root password ${server_root_password}
    mysql -uroot -p${server_root_password} -e (...)
  | COMMAND,
  only_if => "/usr/bin/mysql -u root -e 'show databases;'"
}
\end{minted}
         \caption{Same part of a Chef script rewritten in Puppet}
         \label{fig:puppet-no-smell-ex1}
     \end{subfigure}
     \caption{Inconsistencies in state-of-the-art tools: SLAC reports false positive ``Hard-coded secret'' for script (a); SLIC does not report any security smell for script (b).}
     \label{fig:no-smell-ex1}
\end{figure*}

%To overcome these limitations, we propose 
This paper presents an approach that enables consistent security smell detection across different IaC technologies. We conduct a large-scale empirical study that analyzes security smells on three large datasets containing Ansible, Chef, and Puppet scripts.
%196,756 IaC scripts and 12,281,383 LOC. 
We show that all categories of security smells are identified across all datasets and we identify some smells that might affect many IaC projects.
 To conduct this study, we developed \GLITCH,
%This paper presents \GLITCH, 
a new technology-agnostic framework that enables automated %detection of IaC security smells. \GLITCH\ allows 
polyglot smell detection by transforming IaC scripts into an intermediate representation, on which different security smell detectors can be defined. \GLITCH\ currently supports the detection of nine different security smells and it can analyze scripts written in Puppet, Ansible, or Chef. We compare \GLITCH\ with the state-of-the-art security smell detectors SLIC~\cite{rahman2019seven} and SLAC~\cite{rahman2021security}. The results obtained not only show that \GLITCH\ can reduce the effort of writing security smell analyses for multiple IaC technologies, but also that it has higher precision and recall than the current state-of-the-art tools.
% AQUI OU NA LISTA ABAIXO?
%Moreover, during the development and evaluation of GLITCH, we noticed inconsistencies between available artefacts and their respective research papers. 

%\paragraph{\textbf{Contributions}}
\textbf{Contributions.}
Our main contributions are:
\begin{enumerate*}[label=\textbf{(\arabic*)}]%leftmargin=24pt]
	\item A new intermediate representation that can be used to model IaC scripts and on which security smell detection rules can be defined.
	\item The implementation of a framework called \GLITCH\ that is able to transform IaC scripts written in Ansible, Chef, or Puppet into the new intermediate representation, and that supports the detection of nine security smells. We show that the average precision and recall values of \GLITCH\ are substantially better than the average precision and recall values of state-of-the-art tools.
	\item An empirical study that investigates how frequently security smells occur in IaC scripts. We consider Ansible, Chef, and Puppet scripts. We use three large datasets containing 196,755 IaC scripts and 12,281,251 LOC. We show that all categories of security smells are identified across all datasets and we identify some smells that might affect many IaC projects.
	\item A replication package containing all the datasets used in this work, including three oracle datasets that were manually annotated. We tried to use replication packages from other authors, but they were lacking data. As a result, to the best of our knowledge, our replication package is the first to be complete and available. It is available as a Docker container at
	%{\textbf{\url{https://figshare.com/s/d5283b1cdd1bcee38d85}}}
    {\textbf{\url{https://doi.org/10.6084/m9.figshare.19726603.v2}}}
 
\end{enumerate*}

\medskip
\noindent
\GLITCH\ is open source and available from GitHub:
\begin{center}
\textbf{\url{https://github.com/sr-lab/GLITCH}}
\end{center}

\section{Background and Related Work}
\label{sec:background-related}
%In this work, w
We focus on IaC tools for configuration management of services. The main reason is that the ecosystem around this category of tools is heterogeneous with several technologies widely adopted. 
Guerriero et al.~\cite{guerriero2019adoption} listed four technologies in this category that are adopted by industry experts: Ansible, Chef, Puppet, and Saltstack. Out of these four, three of them had an adoption rate greater than 29\%: Puppet with 29.5\%, Chef with 36.3\%, and Ansible with 52.2\% (note that industry experts can adopt more than one technology simultaneously). To maximize impact of our work, we focus on these three technologies. 
%However, our work can be applied to other tools, even in different categories (e.g., Terraform\footnote{\url{https://www.terraform.io}}).

%The Infrastructure as Code environment has a lot of different technologies able to serve different purposes. Guerriero et al. \cite{guerriero2019adoption} categorized a set of IaC tools into different categories. For instance, the authors state that ``Kubernetes enables orchestration of containers'' and ``Chef and Ansible deal with configuration management of services''. We will focus our work in tools that deal with configuration management of services, namely Ansible, Chef, and Puppet. However, the same techniques could be applied to technologies in other categories, if they share similar concepts (e.g. Terraform\footnote{\url{https://www.terraform.io/}}).

\subsection{Ansible, Chef, and Puppet Scripts}
We provide a brief background on Ansible, Chef, and Puppet scripts. Table \ref{tab:iaccharacteristics} summarizes and compares some relevant characteristics of these technologies. 
There are two types of \emph{configuration setups} for IaC technologies: push and pull. In a push configuration setup, the sysadmin commands a centralized server, able to connect to every node, to provide the configuration to a set of nodes. In a pull configuration setup, each node periodically contacts the server to retrieve the latest configuration for that particular machine. %, and, if it differs from the current one, the new instructions are applied to the node. 
Technologies may require an \emph{additional agent} to be installed in every node. The agent is a program that runs as a background service and is capable of doing the necessary operations in the nodes (e.g., updates). %Usually, pull configuration setups require an agent to be installed. 
Regarding, \emph{syntax}, IaC technologies use different programming languages. 
%Some languages are simpler but have less power, others are more complete and allow to write more complex logic.
Ansible uses YAML, Chef uses Ruby, and Puppet uses a domain-specific language (DSL). Using a programming language like Ruby allows complex programs to be written. However, it may be more difficult to abstract the concepts being represented. 
%Puppet uses a Domain Specific Language based on Ruby.
%\textbf{Execution Order.} 
Ansible and Chef 
%use
encourage a procedural style, which means that scripts follow, in order, a sequence of instructions specified by practitioners. On the other hand, Puppet uses a declarative style, in which practitioners specify the desired state and it is up to the Puppet tool to decide how the state is achieved.
Regarding atomic units and code structure, while Ansible considers the notion of Task as the atomic unit, both Chef and Puppet use the notion of Resource.
In Ansible, configurations are managed using Playbooks, which are decomposed into Plays that define Tasks. In the case of Chef, configurations are defined as Cookbooks, which are decomposed into Recipes specifying Resources. Puppet structures the configurations using Modules that contain configuration files called Manifests. Resources are specified in Classes, which are named blocks used to configure larger chunks of functionality.

\begin{table}
\renewcommand\tabularxcolumn[1]{m{#1}}
\small
\footnotesize
\begin{tabularx}{\linewidth}{XXXX}
\toprule
 \textbf{Characteristic} & \textbf{Ansible}      & \textbf{Chef}         & \textbf{Puppet}        \\ 
\midrule
\textbf{Conf. Setup} & Push & Pull & Pull \\[0.1cm]
\textbf{Add. Agent} & No & Yes & Yes \\[0.1cm]
\textbf{Syntax} & YAML & Ruby & Puppet DSL \\[0.1cm]
\textbf{Exec. Order} & Procedural & Procedural & Declarative \\[0.1cm]
\textbf{Atomic Unit} & Task & Resource & Resource \\[0.1cm]
\textbf{Code \newline Structure} & {\footnotesize Roles \newline - Playbooks \newline - - Plays \newline - - - Tasks} & {\footnotesize Cookbooks \newline - Recipes \newline - - Resources} & {\footnotesize Modules\newline - Manifests\newline - - Classes\newline - - - Resources} \\
\bottomrule
\end{tabularx}
\caption{Summary of Ansible, Chef and Puppet's characteristics.}
\label{tab:iaccharacteristics}
%% JFF: remove vskip later?
\vskip -3em
\end{table}

%\begin{table}
%\renewcommand\tabularxcolumn[1]{m{#1}}
%\small
%\begin{tabularx}{\linewidth}{XXXX}
%\toprule
% \textbf{Characteristic} & \textbf{Ansible}      & \textbf{Chef}         & \textbf{Puppet}        \\ 
%\midrule
%\noalign{\vskip 1mm}
%\textbf{Configuration \newline Setup} & Push & Pull & Pull \\[0.35cm]
%\textbf{Additional \newline Agent} & No & Yes & Yes\\[0.35cm]
%\textbf{Syntax} & YAML & Ruby & Puppet DSL \\[0.2cm]
%%\textbf{Execution \newline Order} & Code order\newline/Procedural & Code order\newline/Procedural & Code order\newline/Declarative \\[0.35cm]
%\textbf{Execution \newline Order} & Procedural & Procedural & Declarative \\[0.35cm]
%%\textbf{Node \newline Variables} & Yes & Yes & Yes \\[0.35cm]
%\textbf{Atomic Unit} & Task & Resource & Resource\\[0.2cm]
%\textbf{Code \newline Structure} & {\footnotesize Roles \newline - Playbooks \newline - - Plays \newline - - - Tasks} & {\footnotesize Cookbooks \newline - Recipes \newline - - Resources} & {\footnotesize Modules\newline - Manifests\newline - - Classes\newline - - - Resources} \\
%\bottomrule
%\end{tabularx}
%\caption{Summary of Ansible, Chef and Puppet's characteristics.}
%\label{tab:iaccharacteristics}
%\end{table}

\subsection{Security Smells in IaC Scripts}
Several catalogs and categories of code smells for IaC scripts have been proposed. Sharma et al.~\cite{sharma2016does} 
%used the knowledge of traditional software engineering and best practices associated with code quality management to create 
created a configuration smells catalog for Puppet scripts
%. The catalog is composed of 
with 13 implementation 
%(e.g. improper alignment of arrows or long statements) 
and 11 design configuration smells. %(e.g. insufficient modularization or duplicate block). 
Schwarz et al.~\cite{schwarz2018code} extended the research done by Sharma et al. by applying the detection of IaC smells to Chef scripts. %The authors also categorized IaC smells as technology agnostic, technology dependent or technology specific, and introduced new code smells from the field of software engineering, such as, Long Resource and Too many Attributes. %Finally, by applying the code smells to repositories with Chef scripts, the results allowed to conclude that “these smells are adequate to be used to investigate the quality of IaC in general”
Rahman and Williams~\cite{rahman2018characterizing} characterized defective IaC scripts by extracting text features from faulty scripts. %The characteristics associated to defective scripts that were found are: file-system operations, infrastructure provisioning, and managing user accounts.
%extracted characteristics of defective IaC scripts from qualitative analysis, which was applied to text features that appeared in faulty scripts. The text features were obtained by applying text mining techniques to convert Puppet scripts into tokens (words in the script). %The characteristics associated to defective scripts that were found are: file-system operations, infrastructure provisioning, and managing user accounts. The authors applied to the tokens two techniques that generate new features capable of being used as input to predictive models: the BOW technique, and the TF-IDF technique. The Random Forest (RF) technique was used to build the models. The datasets to train them were created from crawling open-source projects, generating the features, and manually labeling defective scripts. 10-fold cross validation was used to evaluate the models. These models were able to obtain median F-Measure val- ues between 0.70 and 0.74 depending on the dataset and text feature extraction technique.
Rahman and Williams~\cite{rahman2019source} identified 10 source code properties that correlate with defective IaC scripts.
%focused on source code properties like lines of code, number of attributes, or URL occurrences. They found that using these properties as input to predictive models outperformed the BOW technique from the previous work. The authors also tried to use other techniques to build the models such as Logistic Regression (LR) and Naive Bayes (NB). Depending on the dataset and evaluation metric, different techniques were better, but the Random Forest technique was outperformed in the great majority of the experiments. Rahman and Williams found that the properties with the strongest correlation to a script being defective are the number of lines of code and hard-coded strings.
Rahman et al.~\cite{rahman2020gang} proposed a defect taxonomy for IaC scripts that includes eight categories.
%As in the work to identify security smells [29], Rahman et al. also applied qualitative analysis, but instead of code snippets, the authors used defect-related commits with the goal of identifying defect categories for IaC scripts. They identified 8 categories with configuration data-related defects being the most frequent category, and idempotency being the least frequent one. The authors used the information they got from the qualitative analysis to create empirical rules that automatically identify defect categories in enhanced commit messages (ECMs). ECMs are the combination of commit messages with bug report descriptions that are linked to the commits (e.g. issues). These rules were used to build a tool, called ACID, with a average precision and recall of 0.84 and 0.96, respectively, across all categories
In another work, Rahman et al.~\cite{rahman2020code} identified five development anti-patterns for IaC scripts.
%
%The work listed above is not exclusively focused on security smells. In fact, as 
Focusing on security smells, Rahman et al.~\cite{rahman2019systematic} concluded, after a systematic mapping study with 32 IaC-related publications, that there is a need for more research studies focused on defects and security flaws for IaC. 
Rahman et al.~\cite{rahman2019seven} identified seven types of security smells that are indicative of security weaknesses in Puppet scripts. %They identified 21,201 occurrences of security smells that include 1,326 occurrences of hard-coded passwords. The three languages differ from each other with respect to execution order, perceived
%codebase maintenance, requiring additional agent software installation, style, and syntax. Differ-
%ences in IaC languages along with the need to advance the science of IaC script quality motivate
%us to conduct our research. 
Rahman et al.~\cite{rahman2021security} later replicated this study for Ansible and Chef scripts, identifying two additional security smells.

%{\color{red} TODO: more details about security smells? include description like 3.2 in Rahman et al + 1 or 2 examples}
Like Rahman et al.~\cite{rahman2019seven,rahman2021security}, we also focus on security smells. However, to the best of our knowledge, we are the first to provide a method for enabling consistent security smell detection across different IaC technologies.
%In this work, w
We consider the following nine security smells (we adapt the descriptions by Rahman et al.~\cite{rahman2021security}):
%%% COMPLETAR
\begin{itemize*}[label={}]
    \item \textbf{Admin by default (CWE-250~\cite{mitre2022})}: This smell is the recurring pattern of specifying default users. The smell can violate the ``principle of least privilege'' property~\cite{nist2014}. %, which recommends practitioners to design and implement a system in a manner so that by default the least amount of access necessary is provided to any entity. 
    \item \textbf{Empty password (CWE-258~\cite{mitre2022})}: The smell is the recurring pattern of using a string of length zero for a password. %An empty password is indicative of a weak password.
    \item \textbf{Hard-coded secret (CWE-259, CWE-798~\cite{mitre2022})}: This smell is the recurring pattern of revealing sensitive information, such as user name and passwords in IaC scripts.
    \item \textbf{Unrestricted IP Address (CWE-284~\cite{mitre2022})}: This smell is the recurring pattern of assigning the address 0.0.0.0 for a database server or a cloud service/instance. Binding to the address 0.0.0.0 may cause security concerns as this address can allow connections from every possible network~\cite{mutaf1999defending}.
    \item \textbf{Suspicious comment (CWE-546~\cite{mitre2022})}: This smell is the recurring pattern of putting information in comments about the presence of defects, missing functionality, or weakness of the system (e.g., ``TODO'' and ``FIXME'').
    %. Examples of such comments include putting keywords such as ``TODO'', ``FIXME'', and ``HACK'' in comments, along with putting bug information in comments. 
    \item \textbf{Use of HTTP without SSL/TLS (CWE-319~\cite{mitre2022})}: : This smell is the recurring pattern of using HTTP without the Transport Layer Security (TLS) or Secure Sockets Layer (SSL). Such use makes the communication between two entities less secure~\cite{rescorla2000http}. %, as without SSL/TLS, use of HTTP is susceptible to man-in-the-middle attacks \cite{rescorla2000http}. 
    \item \textbf{No integrity check (CWE-353~\cite{mitre2022})}: This smell is the recurring pattern of downloading content from the Internet and not checking the downloaded content using checksums or gpg signatures.
    \item \textbf{Use of weak cryptography algorithms (CWE-326, CWE-327~\cite{mitre2022})}: This smell is the recurring pattern of using weak cryptography algorithms, namely, MD5 and SHA-1, for encryption purposes.
    \item \textbf{Missing Default in Case Statement (CWE-478~\cite{mitre2022})}: This smell is the recurring pattern of not handling all input combinations when implementing a case conditional logic.
\end{itemize*}

\subsection{Related Work}
%Above, we described related work on catalogs and categories of code smells for IaC scripts. Here, we focus on related work on code quality and security practices, on analysis tools for IaC scripts, and on intermediate representations.

Several studies have been published on code quality and security coding practices for IaC scripts. 
For example, Jiang and Adams~\cite{jiang2015co} conducted an empirical study on the co-evolution of IaC scripts and other software artifacts. They found that the IaC scripts are coupled tightly with the other files in a project. %, especially test files, which implies that testers often need to change infrastructure specifications when making changes to the test framework and tests. 
Hanappi et al.~\cite{hanappi2016asserting} introduced a conceptual framework for asserting reliable convergence in configuration management. 
%Based on a formal definition of configuration scripts and their resources, we utilize state transition graphs to test whether a script makes the system converge to the desired state under different conditions. In our generalized model, configuration actions are partially ordered, often resulting in prohibitively many possible execution orders. To reduce this problem space, we define and analyze a property called preservation, and we show that if preservation holds for all pairs of resources, then convergence holds for the entire configuration. Our implementation builds on Puppet, but the approach is equally applicable to other frameworks like Chef, Ansible, etc. We perform a comprehensive evaluation based on real world Puppet scripts and show the effectiveness of the approach. Our tool is able to detect all idempotence and convergence related issues in a set of existing Puppet scripts with known issues as well as some hitherto undiscovered bugs in a large random sample of scripts.
Van der Bent et al.~\cite{van2018good} proposed a code quality model for Puppet and validated it with experts.
%To this end, we first explore the notion of code quality as it applies to Puppet code by performing a survey among Puppet developers. Second, we develop a measurement model for the maintainability aspect of Puppet code quality. To arrive at this measurement model, we derive appropriate quality metrics from our survey results and from existing software quality models. We implemented the Puppet code quality model in a software analysis tool. We validate our definition of Puppet code quality and the measurement model by a structured interview with Puppet experts and by comparing the tool results with quality judgments of those experts. The validation shows that the measurement model and tool provide quality judgments of Puppet code that closely match the judgments of experts. Also, the experts deem the model appropriate and usable in practice. The Software Improvement Group (SIG) has started using the model in its consultancy practice.

%%% TOOLS
In terms of analysis tools for IaC scripts, 
%there have been proposals for a variety of defects. 
Hanappi et al.~\cite{hanappi2016asserting} propose a tool that detects idempotence and convergence related issues in a set of existing Puppet scripts. Schwarz et al.~\cite{schwarz2018code} picked smells from the catalog proposed by Sharma et al.~\cite{sharma2016does} and convert them into detection rules for Foodcritic, a static code analysis tool designed for Chef. Sotiropoulos et al.~\cite{sotiropoulos2020practical} propose a tool for detecting faults regarding ordering violations and notifiers in Puppet scripts. Lepillet et al.~\cite{lepiller2021analyzing} propose H\"ay\"a, a tool that uses dataflow graph analysis to detect intra-update sniping vulnerabilities in CloudFormation templates.
More relevant for our work are the tools SLIC and SLAC, which are focused on security smells. SLIC, developed by Rahman et al.~\cite{rahman2019seven}, detects seven types of security smells in Puppet scripts and SLAC, developed by Rahman et al.~\cite{rahman2021security}, detects nine types of security smells in Chef scripts and six types in Ansible scripts. Our tool extends the state-of-the-art by providing the first IaC-technology-agnostic framework that can be used to unify tools such as SLIC and SLAC, facilitating the detection of security smells in different IaC technologies. When compared with the security smells supported by SLIC and SLAC, we also identify two additional types of smell in Puppet scripts (\emph{Missing default case statement} and \emph{No integrity check}) and two additional types in Ansible scripts (\emph{Admin by default} and \emph{Use of weak cryptographic algorithms}).

%%% INTERMEDIATE REPRESENTATIONS
Finally, %regarding our use of an intermediate representation, 
%it is worth noting that 
some analysis tools for IaC use intermediate representations \cite{ikeshita2017test,shambaugh2016rehearsal,sotiropoulos2020practical}
%. Namely, these tools use intermediate models 
to describe file-system manipulations done by IaC scripts. %These manipulations include creation and removal of files or directories, creation of links, and renaming a file.
Shambaugh et al.~\cite{shambaugh2016rehearsal} translated IaC scripts to the intermediate representation by mapping types of resources to their filesystem operations. Sotiropoulos et al.~\cite{sotiropoulos2020practical} used system calls executed by each resource in a IaC script to automatically map the resources to the filesystem operations, which were represented in the intermediate language.
To the best of our knowledge, our work is the first that translates scripts of different IaC technologies into an intermediate representation.

%%
%% Intermediate Representation
\section{Intermediate Representation}
\label{sec:intermediate-representation}
\begin{figure}
    \centering
    \includegraphics[width=0.4\textwidth]{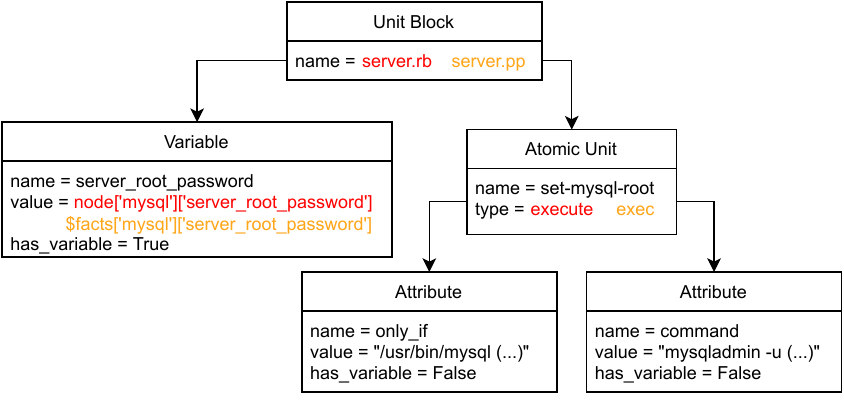}
    \caption{Graph-based representation of the scripts in Figure \ref{fig:no-smell-ex1} using our intermediate representation. In black and red: the representation of the Chef script from Figure \ref{fig:chef-no-smell-ex1}. In black and orange: the representation of the Puppet script from Figure \ref{fig:puppet-no-smell-ex1}.}%
    \label{fig:no-smell-parsing}
\vspace{-1em}
\end{figure}

%Improve title of section.
In order to achieve a technology-agnostic framework, we use an intermediate representation. Our representation is able to capture similar concepts from different IaC technologies, while assuring it is expressive enough to apply analyses that identify security smells. Figure~\ref{fig:formalmodel} describes the abstract syntax of our intermediate representation. We follow an object-oriented approach with a hierarchical structure. As the top-level structure, the intermediate representation can model a  \emph{Project}, a \emph{Module}, or a \emph{Unit block}. Projects represent a generic folder that may contain several modules and unit blocks.
This structure allows us to represent the high-level code structures described in Table~\ref{tab:iaccharacteristics} from Ansible, Chef and Puppet. %The classes that inherit from \textit{CodeElement} save the line and column where the construct appeared to allow the identification of where the smells were in the original code. 
Table \ref{tab:tech-correspondence} shows the relation between the high-level code structures in each IaC technology and the abstract concepts in our intermediate representation. As the table shows, it is possible to find similar structures in the different technologies. \emph{Modules} are the top component from each structure and they agglomerate the scripts necessary to execute a specific functionality. Modules are file system folders, usually with a specific organization (e.g. a role in Ansible usually has a \textit{tasks} and a \textit{vars} folder where, respectively, the tasks and variables for the role are defined). \emph{Unit Blocks} correspond to the IaC scripts themselves or to a group of atomic units. For instance, in Puppet, we can agglomerate resources in classes. %\remove{, and so, the classes are considered unit blocks}.
Finally, \emph{Atomic Units} are the building block of IaC scripts. Atomic units define the system components we want to change and the actions we want to perform on them. 
As shown in Figure~\ref{fig:formalmodel}, unit blocks can have \emph{attribute} definitions, \emph{variable} definitions, and \emph{conditions}. Atomic units have attribute definitions. When values in attribute and variable definitions use variable references, the field \emph{has\_variable} is true. For instance, in Figure \ref{fig:puppet-no-smell-ex1}, the definition of the variable
\mintinline{ruby}{$server_root_password}
%\emph{\$server\_root\_password} 
has as value a reference to the variable
\mintinline{ruby}{$facts['mysql']['server_root_password']},
%\emph{\$facts['mysql']['server\_root\_password']}, 
setting the field \emph{has\_variable} in our intermediate representation to true.
%Although not related to the code structure of the scripts, Table~\ref{tab:tech-correspondence} also refers the concept of switch statements. We focus our attention on these statements since they will be used further ahead in our analyses. Switch statements allow the programmer to define behavior based on the comparison of the value of an expression with multiple values. %and, if the values are equal, to act accordingly to it. 
%These statements also allow to define a default behaviour when neither of the comparisons succeeds. From the three technologies we consider, only Ansible does not have this type of statements.
%
\begin{figure}
\noindent
\begin{minipage}[t]{0.22\textwidth}
\centering
\begin{minted}[fontsize=\scriptsize]{text}
<S> ::= <project> 
      | <module> 
      | <unitblock>

<project> ::= 
    Project {
        name: <str>, 
        modules: <module>*
        blocks: <unitblock>*     
    }

<module> ::= 
    Module {
        name: <str>, 
        blocks: <unitblock>*     
    }
    
<condition> ::=
    ConditionStatement {
        type: IF | SWITCH
        condition: <str>,
        else_statement: <condition>,
        is_default: <bool>
    }

<comment> ::= 
    Comment {
        content: <str>
    }

\end{minted}
\end{minipage}%
\begin{minipage}[t]{0.22\textwidth}
\centering
\begin{minted}[fontsize=\scriptsize]{text}
<unitblock> ::= 
    UnitBlock {
        name: <str>, 
        atomic_units: <atomicunit>*, 
        variables: <variable>*, 
        attributes: <attributes>*
        comments: <comment>*,
        conditions: <condition>*,
        unit_blocks: <unitblock>*
    }

<atomicunit> ::= 
    AtomicUnit {
        name: <str>,
        type: <id>,
        attributes: <attribute>*
    }
    
<attribute> ::= 
    Attribute {
        name: <id>,
        value: <value>,
        has_variable: <bool>
    }

<variable> ::= 
    Variable  {
        name: <id>, 
        value: <value>,
        has_variable: <bool>
    }
\end{minted}
\end{minipage}

\vspace{0.3cm}
\centering
\hbox to \columnwidth{\leaders\hbox to 10pt{\hss - \hss}\hfil}
\vspace{0.1cm}
\begin{minipage}{0.35\textwidth}
\begin{minted}[fontsize=\scriptsize]{text}
<value> ::= <str> | <number> | <bool> | <value>* | <id>
<id> ::= ;sequence of alphanumerics which starts with a letter
<str> ::= "<character>*"    <number> ::= ;integer or double
<bool> ::= True | False
\end{minted}
\end{minipage}
\caption{Abstract syntax of our intermediate representation.}
\label{fig:formalmodel}
\end{figure}
\begin{table}
\caption{Correspondence between the abstract concepts and the concepts in each IaC technology.}
\small
\footnotesize
\begin{tabular}{lccc} \toprule
\textbf{Abstract Concepts} & \textbf{Ansible} & \textbf{Chef} & \textbf{Puppet} \\\midrule
\textbf{Modules} & Roles & Cookbooks & Modules \\
\textbf{Unit Blocks} & Playbooks & Recipes & Manifests, Classes \\ 
\textbf{Atomic Units} & Tasks & Resources & Resources \\
%\textbf{Switch Statements} & - & Case & Case, Selector \\
\bottomrule
\end{tabular}
\label{tab:tech-correspondence}
%% JFF: remove vskip later?
\vskip -2em
\end{table}
Figure~\ref{fig:no-smell-parsing} shows a graph-based visualization of how our intermediate representation models the scripts in Figure~\ref{fig:chef-no-smell-ex1} and Figure~\ref{fig:puppet-no-smell-ex1}.
%Similar scripts from different technologies will result in similar structures in our representation. This means that if the same rules for smell detection are applied to these scripts, \GLITCH\ will be consistent in the reported security smells. 

%%
%% GLITCH
\section{Security Smell Detection}
\label{sec:glitch}

In Table \ref{tab:security-rules}, we define the rules used by \GLITCH\ to detect security smells. The formalism used to define rules is similar to the one used by SLIC~\cite{rahman2019seven} and SLAC~\cite{rahman2021security}. The functions \textit{isAttribute}, \textit{isVariable}, \textit{isComment}, \textit{isAtomicUnit}, and \textit{isConditionStatement} verify the type of instance being analyzed (e.g., if the node $x$ is an Attribute node, \textit{isAttribute(x)} is true). Each node in our representation is referred by the variable \textit{x}. We traverse the nodes using a depth-first search (DFS). We start in the initial node (a Project, a Module or a Unit Block) and then we execute the DFS considering each collection inside the node as its children. Each node may have more than one security smell, and so every rule is applied, even if a smell was already identified for that node. Previous nodes have no influence in the analyses of other nodes. The function \textit{hasDownload} goes through a list of attributes and verifies if for at least one of them \textit{isDownload(x.value)} is true. The same goes for the function \textit{hasChecksum} but instead of using \textit{isDownload}, it uses \textit{isChecksum}. The function \textit{isDefault} is a recursive function that returns true if a default branch is found in the case statement, and false otherwise. The remaining functions are defined in Table \ref{tab:functions-glitch}. These functions verify if any of the string patterns described are present in the values they receive.

The \GLITCH\ framework allows the definition of different configurations to identify security smells. These configurations change the keywords in the (disjunctive) string patterns for each function defined in Table \ref{tab:functions-glitch}. In the table, we describe the configuration used by the improved version of \GLITCH\ to which we will refer in Section~\ref{sec:evaluation}. 
%The existence of c
Configurations allow users to tweak the tool to best suit the needs of the IaC developers and to better adapt to each IaC technology. 
%\remove{If we consider the example in Figure \ref{fig:no-smell-parsing}, we can observe that if we apply the rule \textit{Hard-coded secret} to the variable node, \GLITCH\ will not trigger a security smell for neither technology, since the \textit{has\_variable} flag is set to true.}
%
\GLITCH\ is implemented in Python and it currently supports the analysis of Ansible, Chef, and Puppet scripts. Our implementation transforms the original scripts into our intermediate representation and then attempts to detect security smells as described above. To parse the Ansible scripts we used the \textit{ruamel.yaml} package\footnote{\url{https://pypi.org/project/ruamel.yaml/}} for Python. The Chef scripts were parsed using Ripper,\footnote{\url{https://github.com/ruby/ruby/tree/master/ext/ripper}} a script parser for Ruby. We developed a parser for Ripper's output using a package called \textit{ply}.\footnote{\url{https://github.com/dabeaz/ply}} Finally, for Puppet scripts, we developed our own parser\footnote{\url{https://github.com/Nfsaavedra/puppetparser}} using the same \textit{ply} package. We decided to develop our parser since we did not find any other good options to parse Puppet DSL in Python.

\begin{table*}
\caption{Rules to detect security smells used by \GLITCH.}
\label{tab:security-rules}
\scriptsize
\begin{tabularx}{\linewidth}{r L{0.8}}
\hline
\textbf{Smell Name} & \textbf{Rule}         \\ \hline
Admin by default & $(isAttribute(x) \lor isVariable(x)) \land (isUser(x.name) \lor isRole(x.name)) \land isAdmin(x.value) \land \neg x.has\_variable$\\
Empty password & $(isAttribute(x) \lor isVariable(x)) \land isPassword(x.name) \land length(x.value)==0$\\
Hard-coded secret & $(isAttribute(x) \lor isVariable(x)) \land (isPassword(x.name) \lor isSecret(x.name) \lor isUser(x.name)) \land \neg x.has\_variable$\\
Invalid IP address binding & $(isAttribute(x) \lor isVariable(x)) \land isInvalidBind(x.value)$\\
Suspicious comment & $isComment(x) \land hasWrongWords(x.content)$\\ 
Use of HTTP without TLS & $(isAttribute(x) \lor isVariable(x)) \land isURL(x.value) \land hasHTTP(x.value) \land \neg hasHTTPWhiteList(x.value)$\\
No integrity check & $(isAtomicUnit(x) \land hasDownload(x.attributes) \land \neg hasChecksum(x.attributes)) \lor ((isAttribute(x) \lor isVariable(x)) \land isCheckSum(x.name) \land (x.value == "no" \lor x.value == "false"))$\\
Use of weak crypto alg. & $(isAttribute(x) \lor isVariable(x)) \land isWeakCrypt(x.value) \land \neg hasWeakCryptWhiteList(x.name) \land \neg hasWeakCryptWhiteList(x.value)$\\
Missing default case statement & $isConditionStatement(x) \land x.is\_default == False \land \neg isDefault(x.else\_statement)$\\ \hline
\end{tabularx}
\end{table*}

\begin{table}
%\caption{String patterns used for the functions in the \GLITCH's rules. This configuration is the one used by the improved version of \GLITCH.}
\caption{String patterns used in the \GLITCH's rules. These are configurable. The configuration shown is the one used by the improved version of \GLITCH.}
\label{tab:functions-glitch}
\scriptsize
\begin{tabularx}{\linewidth}{r L{0.8}}
\hline
\textbf{Function} & \textbf{String Pattern}         \\ \hline
isUser() & "user", "uname", "username", "login", "userid", "loginid" (...)\\
isRole() & (the config is empty for this function)\\
isAdmin() & "admin", "root"\\
isPassword() & "pass", "pwd", "password", "passwd", "passno", "pass-no" (...)\\
isSecret() & "auth\_token", "authetication\_token", "secret", "ssh\_key" (...)\\
isInvalidBind() & "0.0.0.0"\\
hasWrongWords() & "bug", "debug", "todo", "hack", "solve", "fixme" (...)\\
hasHTTP() & "http"\\
hasHTTPWhiteList() & "localhost", "127.0.0.1" \\
isDownload() & "(http|https|www).*iso\$", "(http|https|www).*tar.gz\$" (...)\\
isCheckSum() & "gpg", "checksum" \\
isWeakCrypt() & "md5", "sha1", "arcfour" \\
hasWeakCryptWhiteList() & "checksum" \\ \hline
\end{tabularx}
\vspace{-2em}
\end{table}

%%
%% Evaluation
\section{Evaluation}
\label{sec:evaluation}
This section describes the evaluation of \GLITCH. 
%JFF: se conseguirmos, voltar a colocar isto
%We start by presenting the research questions (Section~\ref{sec:eval:rqs}). We then describe how we constructed the datasets used in the evaluation (Section~\ref{sec:eval:datasets}). Using oracle datasets, we compare the accuracy of \GLITCH\ with the accuracy of the tools SLIC~\cite{rahman2019seven} and SLAC~\cite{rahman2021security} (Section~\ref{sec:eval:accuracy}). Finally, we use \GLITCH\ to perform an empirical study of security smells in Ansible, Chef, and Puppet scripts (Section~\ref{sec:eval:smells-frequency}).

\subsection{Research Questions}
\label{sec:eval:rqs}
% As respostas a estas RQs poderão ser dadas ou no final das seccoes da S5 ou na S6
%We aim to answer the following research questions:
                                                        
\begin{itemize}%[leftmargin=24pt]                                      
\item[\textbf{RQ1}.] {\textbf{[Abstraction]}} Can our intermediate representation model IaC scripts and support automated detection of security smells?

%%In this first research question, w
%We are interested in determining whether our intermediate representation is capable of abstracting relevant information from IaC scripts written in different IaC languages, so that one can define security smell detectors on it.

\item[\textbf{RQ2}.] {\textbf{[Accuracy and Performance]}} How does \GLITCH\ compare with existing state-of-art tools for detecting security smells in terms of accuracy and performance?

%%In this second research question, w
%We are interested in comparing the accuracy and performance of \GLITCH\ with the accuracy of existing tools, such as SLIC~\cite{rahman2019seven} and SLAC~\cite{rahman2021security}.

\item[\textbf{RQ3}.] {\textbf{[Frequency]}} How frequently do security smells occur in IaC scripts?

%%In this third research question, w
%We are interested in characterizing how frequently security smells are present in IaC scripts. This research question was addressed by Rahman et al.~\cite{rahman2019seven} for Puppet and by Rahman et al.~\cite{rahman2021security} for Ansible and Chef. They used different tools for answering this question. Here, we want to use \GLITCH\ and to investigate whether there are any noticeable differences.
\end{itemize}

\subsection{Datasets}
\label{sec:eval:datasets}
% \begin{table*}
% \small
% \caption{Attributes of IaC Datasets.}
% \label{tab:datasets-metrics}
% \centering
% \begin{tabular}[t]{lrrrrrrr}
% \toprule
% %& \multicolumn{1}{c}{\textbf{Ansible}} & \multicolumn{1}{c}{\textbf{Chef}} & \multicolumn{4}{c}{\textbf{Puppet}} \\
% &  & & \multicolumn{4}{c}{\textbf{Puppet}} \\
% \cmidrule(lr){4-7}
% \textbf{Attribute} & \textbf{Ansible} & \textbf{Chef} & \textbf{GH} & \textbf{MOZ} & \textbf{OST} & \textbf{WIK} \\
% \midrule
% Repository count & 681 & 439 & 219 & 2 & 61 & 11 \\
% Total IaC scripts & 108,510 & 70,939 & 10,009 & 1613 & 2,840 & 2,845\\
% Total LOC (IaC scripts) & 5,180,879 & 6,071,035 & 610,122 & 66,367 & 217,843 & 135,137\\
% \bottomrule
% \end{tabular}
% \end{table*}
\begin{table*}
\footnotesize
\small
\begin{minipage}[b]{0.6\textwidth}
\centering
\caption{Attributes of IaC Datasets.}
\label{tab:datasets-metrics}
\begin{tabular}[t]{lrrrrrrr}
\toprule
%& \multicolumn{1}{c}{\textbf{Ansible}} & \multicolumn{1}{c}{\textbf{Chef}} & \multicolumn{4}{c}{\textbf{Puppet}} \\
&  & & \multicolumn{4}{c}{\textbf{Puppet}} \\
\cmidrule(lr){4-7}
\textbf{Attribute} & \textbf{Ansible} & \textbf{Chef} & \textbf{GH} & \textbf{MOZ} & \textbf{OST} & \textbf{WIK} \\
\midrule
Repository count & 681 & 439 & 219 & 2 & 61 & 11 \\
Total IaC scripts & 108,509 & 70,939 & 10,009 & 1613 & 2,840 & 2,845\\
Total LOC (IaC scripts) & 5,180,747 & 6,071,035 & 610,122 & 66,367 & 217,843 & 135,137\\
\bottomrule
\end{tabular}
\end{minipage}%
\begin{minipage}[b]{0.40\textwidth}
\centering
\caption{Attributes of Oracle Datasets.}
\label{tab:oracle-datasets-metrics}
\begin{tabular}[t]{lrrrr}
\toprule
\textbf{Attribute} & \textbf{Ansible} & \textbf{Chef} & \textbf{Puppet} \\
\midrule
Total IaC scripts & 81 & 80 & 80\\
Total LOC (IaC scripts) & 4,185 & 4,630 & 4,367\\
\bottomrule
\end{tabular}
\end{minipage}
\end{table*}

This section describes how we constructed the datasets used for our evaluation. Since we consider Ansible, Chef, and Puppet scripts, our first step was to attempt to obtain the same datasets as used in the studies involving SLIC and SLAC~\cite{rahman2019seven, rahman2021security}. We got hold of the publicly available datasets\footnote{\url{https://doi.org/10.6084/m9.figshare.8085755}} and Docker image\footnote{\url{https://hub.docker.com/repository/docker/akondrahman/slic_ansible}}, and we observed that only the oracle for Ansible was available. We thus contacted the first author of the studies mentioned above, who very kindly shared with us a Puppet dataset almost identical to the one used in the empirical study using SLIC (there were small differences in the number of Puppet scripts contained in the dataset). 
We constructed oracle datasets for Chef and Puppet as these oracle datasets were not available as part of Rahman et al.'s replication packages. We further contacted the first author about the availability of the oracle datasets 
%used in Rahman et al., 
and learned that these datasets reside in computing clusters to which the first author no longer has access to.
%
%He also informed us that the Puppet oracle was not available, since it was stored in an AWS instance that no longer exists; and that none of the Chef datasets are available, since they were in a laptop that is no longer accessible. He also shared a larger Ansible dataset, but it was not as complete as the one described in the evaluation of SLAC~\cite{rahman2021security}.
%
Given this, we decided to reuse their oracle for Ansible and the Puppet dataset, and to construct new oracles for Chef and Puppet, and new IaC datasets for Ansible and Chef.

\subsubsection{IaC datasets}
To perform an empirical study of security smells in Ansible, Chef,
and Puppet scripts, we require three datasets of IaC scripts, one for each technology.
As mentioned above, we reused Rahman et al.'s Puppet dataset~\cite{rahman2019seven}, which is composed of four different sub-datasets. Three datasets are constructed using repositories collected from three organizations: Mozilla (MOZ), Openstack (OST), and
Wikimedia (WIK). The fourth dataset is constructed from repositories hosted on GitHub (GH). 
%To assess the prevalence of the identified smells and increase generalizability of our findings, we include repositories from Github, as companies tend to host their popular OSS projects on GitHub [41] [42].

For Ansible and Chef, we created two new datasets by selecting OSS repositories from GitHub.
As described in previous research~\cite{munaiah2017curating}, OSS repositories need to be curated. We apply the same criteria that Rahman et al.~\cite{rahman2019seven} used to construct their Puppet sub-datasets extracted from GitHub (except that we consider all the available repositories created between 2012 and 2022):

\begin{itemize*}[label={}]
    \item \textbf{Criterion 1}:
    At least 11\% of the files belonging to the repository must be IaC scripts. This follows from a Jiang and Adams's study~\cite{jiang2015co}, where it was observed that in OSS repositories, a median of 11\% of the files are IaC scripts. The rationale is to collect repositories that contain sufficient amount of IaC scripts for analysis.
    \item \textbf{Criterion 2}:
    The repository is not a clone. 
    \item \textbf{Criterion 3}:
    The repository must have at least two commits per month. This is based on Munaiah et al.~\cite{munaiah2017curating}, who used the threshold of at least two commits per month to determine which repositories have enough software development activity.
    \item \textbf{Criterion 4}:
    The repository has at least 10 contributors. Similar to Rahman et al.~\cite{rahman2019seven}, we assume that this criterion may help us to filter out irrelevant repositories.
\end{itemize*}

Table~\ref{tab:datasets-metrics} presents the number of repositories, the number of IaC scripts, and the number of LOC in the three IaC datasets. The Ansible dataset was constructed from 681 repositories and contains 108,509 Ansible scripts (5,180,747 LOC).
The Chef dataset was constructed from 439 repositories and contains 70,939 Chef scripts (6,071,035 LOC). The Puppet dataset was constructed from 293 repositories and contains 17,307 Puppet scripts (1,029,469 LOC).
When considering the three IaC datasets as a whole, there are 1413 repositories with 196,755 IaC scripts. In total, there are 12,281,251 LOC.

\subsubsection{Oracles}
% \begin{table}
% \small
% \caption{Attributes of Oracle Datasets.}
% \label{tab:oracle-datasets-metrics}
% \centering
% \begin{tabular}[t]{lrrrr}
% \toprule
% \textbf{Attribute} & \textbf{Ansible} & \textbf{Chef} & \textbf{Puppet} \\
% \midrule
% Total IaC scripts & 81 & 80 & 80\\
% Total LOC (IaC scripts) & 4,185 & 4,630 & 4,367\\
% \bottomrule
% \end{tabular}
% \end{table}
To determine the accuracy of \GLITCH\ and to compare it with other tools, we require three oracle datasets, one for each IaC technology considered. In what follows, we describe how we selected the IaC scripts included in each oracle and how we annotated the datasets.

\paragraph{File collection.}
For the Ansible oracle, we reused Rahman et al.'s oracle~\cite{rahman2021security}, which contains 81 IaC scripts. 
We constructed new oracle datasets for Chef and Puppet. To ensure that the size of the three oracles was similar, based on the size of the Ansible oracle dataset, we decided to create oracles with exactly 80 IaC scripts. To select the files, we wrote a Python script that kept selecting a random file from the respective IaC dataset described in the previous subsection
while the desired size was not achieved.
For each file, we ran \GLITCH\ and either SLAC (if the file was a Chef script) or SLIC (if the file was a Puppet script). We kept track of the number of security smells reported and their respective categories. If, after analyzing a file, the file contained a smell of a category that up to that point had less than 5 reports, then the file was included in the oracle dataset.
% After the minimum number of reports for each smell was achieved, the remaining files were added to the dataset without restrictions.
Table~\ref{tab:oracle-datasets-metrics} presents the number of IaC scripts and the number of LOC in the three oracle datasets.

\paragraph{Annotating the oracle datasets.}
After collecting the scripts that make the oracle datasets, we manually annotated them, identifying security smells. 
Despite the use of analysis tools in the file selection process described above, we guaranteed that the location of the security smells was not disclosed. In other words, at the annotation stage we only had access to the files, but not the reports. We did this to reduce bias in the annotation process.
The Ansible oracle dataset was already annotated, but since the numbers of smell occurrences did not match the numbers reported in Rahman et al.'s study~\cite{rahman2021security}, we decided to reannotate the dataset.
To annotate the oracle datasets, we used closed coding~
\cite{saldana2021coding}, where three raters identified security smells and their agreement was checked.
In total, there were seven raters involved. One of the raters was the first author. For each of the three IaC technologies, we recruited two postgraduate students who had experience with IaC and/or cybersecurity. They were given access to: the 80 files in the oracle datasets, a general description of the IaC technology, and a description of the nine security smells considered. For each report, raters identified the name of the file, the category of the security smell, and the line where it occurs; they collated this information in a CSV file.

We then manually inspected the three CSV files produced for each oracle dataset, and we decided to keep only the classifications where at least two raters agreed. 
Table \ref{tab:glitch-oracle-agreement} shows the agreement distribution for each dataset. 
We only consider the lines of code
%We defined our subjects as the lines of code 
where at least one rater identified a smell.
The percentage values shown are for the cases where there was
%Each cell has the percentage of subjects on which there was
no agreement, two raters agreed, or all the raters agreed. 
%TODO: reescrever subjects acima?
When a rater did not identify a smell identified by other rater, we considered the label ``none'' to be attributed.
%When other raters did not identify a smell on the same line of code, we considered the label ``none'' to be attributed.
The results on the table demonstrate that at least two raters agreed on the great majority of subjects: 99.1\% in Ansible, 93.9\% in Chef, and 95.1\% in Puppet. We calculated the agreement distribution instead of other statistics, such as Cohen's Kappa or Krippendorff's alpha, since these statistics consider the probability of chance agreement. 
We argue that, since our annotation task includes finding the smells in the scripts, the likelihood of chance agreement is significantly reduced.
%Our annotation task includes finding the smells in the scripts which makes the probability of chance agreement negligible. 
%We could also have calculated the percent agreement (which ignores the probability of chance agreement), however this statistic strongly penalizes cases where 2 out of 3 raters agree, which would not fairly represent how good our classifications are when considering at least 2 raters.}
%
\begin{table}
\small
\caption{Agreement distribution for the oracle datasets (\%).} % of subjects).}
\label{tab:glitch-oracle-agreement}
\centering
\begin{tabular}[t]{lrrr}
\toprule
& \textbf{Ansible} & \textbf{Chef} & \textbf{Puppet} \\\midrule
No agreement & 0.9 & 6.1 & 4.9\\
2 raters agreed & 86.4 & 73.6 & 78.6\\
3 raters agreed & 12.7 & 20.3 & 16.5\\
\bottomrule
\end{tabular}
\end{table}
% MORE STATISTICS? PERCENTAGEM DE AGREEMENT ENTRE OS 3? ETC.
After this process, we obtained: an oracle of 44 Ansible security smells categorized as shown in Table~\ref{tab:glitch-slac-oracle-ansible} and with 69 files with no smells; an oracle of 105 Chef security smells categorized as shown in Table~\ref{tab:glitch-slac-oracle-chef} and with 43 files with no smells; and an oracle of 65 Puppet security smells categorized as shown in Table~\ref{tab:glitch-slic-oracle-puppet} and with 52 files with no smells.

%Se pelo menos dois concordavam, considerei smell. Existe uma tag de "none" para quando não se encontra smells, que causava problemas neste sistema no entanto:
%Eu - none
%Bernardo - No integrity check
%David - Hard-coded secret
%Neste caso, em que não há concordância de pelo menos 2 para nenhum smell, considerei sempre none, mesmo não havendo 2 a concordar.

%Há um ou dois casos entre todos os datasets em que considerei smell mesmo não havendo concordância porque eram erros que se via que foi distração e que eram óbvios. Mas os datasets finais têm uma coluna AGREEMENT, onde diz tudo isto
%￼

\subsection{Accuracy of \GLITCH}
\label{sec:eval:accuracy}
To determine the accuracy of \GLITCH, we ran it for the oracle datasets. We also ran SLIC for the Puppet oracle dataset and SLAC for the other two oracle datasets.
We measured precision and recall of each tool. 
%Precision refers to the fraction of correctly identified smells among the total identified security smells, as determined by each tool. Recall refers to the fraction of correctly identified smells that have been retrieved by each tool over the total amount of security smells. 
Since it is easy to configure \GLITCH\ (see Section~\ref{sec:glitch}), we used two versions of \GLITCH\ for each oracle dataset: one version was configured to behave similarly to SLIC (or SLAC), and the other was an improved version. As described in Section~\ref{sec:glitch}, the difference between the two versions is on the keywords for each function in Table \ref{tab:functions-glitch}: one uses the keywords used by SLIC (or SLAC) and the other configuration was tweaked by us. In the tables below, we use the headers GLITCH (SLIC) and GLITCH (SLAC) to refer to GLITCH configured to behave similarly to SLIC and SLAC, respectively. The header GLITCH refers to the improved version of GLITCH that uses the rules shown in Table~\ref{tab:functions-glitch}.

Tables~\ref{tab:glitch-slac-oracle-ansible},~\ref{tab:glitch-slac-oracle-chef},~and~\ref{tab:glitch-slic-oracle-puppet} report the accuracy results for Ansible, Chef, and Puppet, respectively. We use N/I to denote that the detection of a certain smell is not implemented (e.g., SLAC does not detect the smell \emph{Admin by default} for Ansible scripts); N/A to denote that a certain smell cannot occur (e.g., Ansible does not have switch statements, so the smell \emph{Missing default case statement} does not apply); and N/D to denote that the tool does not report any security smell or to denote that there are no occurrences of a given smell (see, for example, the recall value of \GLITCH\ for the \emph{Use of weak crypto algorithm} in Table~\ref{tab:glitch-slac-oracle-ansible}). 
To facilitate comparison between tools and IaC technologies, we decided to keep all the rows in these tables, even when there are no smell occurrences or when its detection is not implemented.

\subsubsection{Accuracy results for the Ansible oracle dataset}
As shown in Table~\ref{tab:glitch-slac-oracle-ansible}, \GLITCH\ configured to behave similarly to SLAC has the same precision and recall as SLAC (same average). %There is a discrepancy in the precision for the smell \emph{Hard-coded secret} (32\% vs 33\%) that also creates a small discrepancy in the recall values for \emph{No Smell}. This happens because one of the Ansible scripts %openstack-ansible-ops-elk_metrics_6x-roles-elasticsearch-meta-main.yml, 
%is a metadata file with no security smells identified in the oracle. While \GLITCH\ does not parse it (thus not reporting any smell for it), SLAC reports a false positive of type \emph{Hard-coded secret}.
There is a small discrepancy in the recall values for \emph{No Smell}. This happens because SLAC detects one \emph{No integrity check} smell in an Ansible script  %browbeat-ansible-install-roles-epel-defaults-main.yml,
where no smells should be detected. The difference between both tools is that GLITCH enforces detection of \emph{No integrity check} smells only on \emph{Atomic Unit} nodes, while SLAC ignores the type of node, which leads SLAC to detect this type of smell in the definition of a variable.

Regarding the improved version of \GLITCH, the average precision improves from 67\% to 77\% and recall improves from 79\% to 87\%. There are also improvements regarding files with no smells. We can also see that it supports the smell \emph{Admin by default} with perfect precision and recall. \GLITCH\ keeps the values of precision and recall when they were already 100\%. It also improves the precision for \emph{Hard-coded secret} by 10 percentage points (from 32\% to 42\%); for \emph{Suspicious comment} by 8 percentage points (from 67\% to 75\%); and for \emph{Use of HTTP without TLS} by 24 percentage points (from 71\% to 95\%). Recall for \emph{Suspicious comment} improved from 67\% to 100\%.
The only case where improvements do not occur is for the smell \emph{No integrity check}, where the single occurrence is not detected (note that SLAC did not detect it either). This happens because the occurrence of this smell is regarding a URL referring to an YAML file, which \GLITCH\ does not consider (i.e., the string pattern \textit{isDownload()} shown in Table~\ref{tab:functions-glitch} does not contain URLs that end with \texttt{.yml}). Finally, the worst precision value is for the smell \emph{Hard-coded secret} (42\%). This happens mainly because the string patterns \textit{isSecret()}, \textit{isPassword()}, and \textit{isUser()} are the ones with more possibilities, thus increasing the probability of having false positives. Some of the possibilities are keywords such as ``user'', which result in a higher number of false positives.

\subsubsection{Accuracy results for the Chef oracle dataset}
Table~\ref{tab:glitch-slac-oracle-chef} shows that when \GLITCH\ is configured to behave similarly to SLAC, it actually obtains better results than SLAC: 
the average precision improves 28 percentage points (from 49\% to 77\%) and the average recall improves 16 percentage points (from 60\% to 76\%). There are also improvements regarding files with no smells.
Contributing to these improvements is the substantial increase in precision for the smells \emph{Empty password} and \emph{No integrity check}. Regarding the first smell, this is because SLAC wrongly treats variables as empty values;
%\GLITCH\ seems to deal much better with variables as values than SLAC; 
regarding the second, \GLITCH\ searches for links in the values of variables and attributes, while SLAC is searching for links on a line-by-line basis. 
%\review{It is worth noting that there are no false positives with respect to the smell \emph{Admin by default}.}
%% JFF: comentamos para poupar espaco
%Also, the recall value for the smell \emph{No integrity check} decreases. This happens because SLAC identified one additional true positive. The reason for this is because SLAC searches for links on a line-by-line basis, while \GLITCH\ checks whether an attribute or variable is a link. Since in one of the examples there is an attribute where the value \emph{contains} a link as part of the value (but also contains other data), \GLITCH\ is unable to detect, but SLAC is.

When compared to \GLITCH\ configured to behave similarly to SLAC, the improved version 
maintains the average precision
%improves the average precision by only 1\% (from 76\% to 77\%) 
and increases the average recall by 10 percentage points (76\% to 86\%). When compared to SLAC, the results for all smells improve, except for \emph{Invalid IP address binding} and \emph{Use of HTTP without TLS}, where the results are the same, and for \emph{Suspicious comment}, where the precision decreases. 
%Similarly to what happened for the smell \emph{Hard-coded secret} when analyzing the Ansible oracle dataset, t
This decrease in precision is because \GLITCH\ uses a larger set of keywords (this is similar to what caused the low precision for the smell \emph{Hard-coded secret} when analyzing the Ansible oracle dataset). This is also why the worst precision value is for the smell \emph{Hard-coded secret}. The worst recall value is for the smell \emph{Admin by default} (41\%). This happens because there are some scripts in the dataset that configure the execution of MySQL commands. The commands executed as root, such as the following, were considered by the raters as a security smell: ~\verb|cmd = "mysql -uroot ..."|.
However, for this smell, \GLITCH\ only considers the value of attributes or variables that define users (e.g. \texttt{user: root}).

%%%%%%%%%%%%%%%%%%%%%%%%%%%%%%%%%%%%%%%%%%%
% Chef Results
%%%%%%%%%%%%%%%%%%%%%%%%%%%%%%%%%%%%%%%%%%%
% GLITCH configured to behave similarly to SLAC
% - presents better results: average precision improves 27% (from 49% to 76%) and average recall improved 13% (from 61% to 74%). TODO: EXPLAIN WHY (no Slack)
% - The precision and recall for No Smell improved.
% - Admin by default: no false positives
% - Empty password and No integrity check: precision improves to 100%
% - No integrity check: recall gets worse. WHY? (no Slack)
%
%
% Improved version of GLITCH
% - When compared to GLITCH configured to behave similarly to SLAC, the improved version improved the average precision by 1% (from 76% to 77%) and the average recall by 10% (74% to 84%).
% - When compared to SLAC, the results for all smells improved, except Suspicious comment: WHY? (no Slack)
% - Empty password and No integrity check: precision improves to 100% and recall improves considerably
% The worst precision is for Hard-coded secret: (Same reason as in Ansible)
% The worst recall is for Admin by default: This is related to a particularity of the dataset. There are some scripts in the dataset that configure the execution of MySQL commands. The commands executed as root were considered by practitioners as a smell. However, when we identify this type of smell, we check if the name of the attribute/variable is related to the definition of a user, and, in the cases we get false negatives, they are not.

% cmd = "mysql -uroot -p#{get_config!('password','mysql-root','os')} -e"

\subsubsection{Accuracy results for the Puppet oracle dataset}
Similar to what was described above, Table~\ref{tab:glitch-slic-oracle-puppet} shows that when \GLITCH\ is configured to behave similarly to SLIC, it also obtains better results than SLIC:  the average precision improves 8 percentage points (from 60\% to 68\%) and the average recall improves 10 percentage points (from 72\% to 82\%). 
Contributing to this is the fact that \GLITCH\ detects smells of type \emph{Missing default case statement} with a high precision. Also, the precision for the smell \emph{Empty password} is noticeable higher (\GLITCH\ reports no false positives). This is because \GLITCH\ seems to deal better with variables. There are also improvements regarding files with no smells.

When compared to \GLITCH\ configured to behave similarly to SLIC, the improved version maintains the average precision and improves the average recall by 3 percentage points (82\% to 85\%). 
The precision and recall for \emph{No Smell} decreased 1 and 6 percentage points, respectively.
We can see that for the smell \emph{Admin by default} many more true positives are identified, but there are some false positives. There were no reports for the smell \emph{No integrity check}. Precision and recall improved or remained the same for all the smells, except for \emph{Suspicious comment}. Similar to what happened with the Chef oracle dataset, the precision values for the smells \emph{Hard-coded secret} and \emph{Suspicious comment} are low due to the use of more keywords.

\begin{table*}
\small
\footnotesize
\caption{\GLITCH\ vs SLAC: Accuracy for the Ansible Oracle Datasets \scriptsize{(N/I - Not implemented, N/A - Not applicable, N/D - No data)}}
\label{tab:glitch-slac-oracle-ansible}
\centering
\begin{tabular}[t]{lrrrrrrr}
\toprule
& \multicolumn{7}{c}{\textbf{Original Oracle}} \\
\cmidrule(lr){2-8}
& & \multicolumn{2}{c}{\textbf{SLAC}} & \multicolumn{2}{c}{\textbf{\GLITCH\hspace{0.1cm}(SLAC)}} & \multicolumn{2}{c}{\textbf{\GLITCH}}\\
\cmidrule(lr){3-4} \cmidrule(lr){5-6} \cmidrule(lr){7-8}
\textbf{Smell Name} & \textbf{Occurr.} & \textbf{Precision} & \textbf{Recall} & \textbf{Precision} & \textbf{Recall} & \textbf{Precision} & \textbf{Recall}\\
\midrule
Admin by default & 7 & N/I & N/I & N/I & N/I & 1.00 & 1.00  \\
\rowcolor{Gray}
Empty password   & 1 & 1.00 & 1.00 & 1.00 & 1.00 & 1.00 & 1.00 \\
Hard-coded secret   & 8 & 0.32 & 1.00 & 0.32 & 1.00 & 0.42 & 1.00 \\
\rowcolor{Gray}
Invalid IP address binding   & 1 & 1.00 & 1.00 & 1.00 & 1.00 & 1.00 & 1.00\\
Suspicious comment   & 6 & 0.67 & 0.67 & 0.67 & 0.67 & 0.75 & 1.00\\
\rowcolor{Gray}
Use of HTTP without TLS   & 20 & 0.71 & 1.00 & 0.71 & 1.00 & 0.95 & 1.00\\
No integrity check       & 1 & 0.00 & 0.00 & 0.00 & 0.00 & 0.00 & 0.00\\
\rowcolor{Gray}
Use of weak crypto alg.   & 0 & N/I & N/I & N/I & N/I & N/D & N/D\\
Missing default case statement & 0 & N/A & N/A & N/A & N/A & N/A & N/A\\
\midrule
No smell   & 69 & 0.98 & 0.87 & 0.98 & 0.88 & 1.00 & 0.94\\
\midrule
\textbf{Average}   && 0.67 & 0.79 & 0.67 & 0.79 & 0.77 & 0.87\\
\bottomrule
\end{tabular}
%\caption*{\scriptsize{(N/I - Not implemented, N/A - Not applicable, N/D - No data)}}
\end{table*}

\begin{table*}
\small
\footnotesize
\caption{\GLITCH\ vs SLAC: Accuracy for the Chef Oracle Datasets \scriptsize{(N/I - Not implemented, N/A - Not applicable, N/D - No data)}}
\label{tab:glitch-slac-oracle-chef}
\centering
\begin{tabular}[t]{lrrrrrrr}
\toprule
& \multicolumn{7}{c}{\textbf{Original Oracle}} \\
\cmidrule(lr){2-8}
& & \multicolumn{2}{c}{\textbf{SLAC}} & \multicolumn{2}{c}{\textbf{\GLITCH\hspace{0.1cm}(SLAC)}} & \multicolumn{2}{c}{\textbf{\GLITCH}}\\
\cmidrule(lr){3-4} \cmidrule(lr){5-6} \cmidrule(lr){7-8}
\textbf{Smell Name} & \textbf{Occurr.} & \textbf{Precision} & \textbf{Recall} & \textbf{Precision} & \textbf{Recall} & \textbf{Precision} & \textbf{Recall}\\
\midrule
Admin by default & 37 & 0.00 & 0.00 & N/D & 0.00 & 0.94 & 0.41  \\
\rowcolor{Gray}
Empty password & 4 & 0.00 & 0.00 & 1.00 & 0.75 & 1.00 & 0.75 \\
Hard-coded secret & 13 & 0.13 & 0.54 & 0.20 & 0.69 & 0.20 & 0.69 \\
\rowcolor{Gray}
Invalid IP address binding & 7 & 1.00 & 1.00 & 1.00 & 1.00 & 1.00 & 1.00\\
Suspicious comment & 4 & 0.80 & 1.00 & 0.80 & 1.00 & 0.40 & 1.00\\
\rowcolor{Gray}
Use of HTTP without TLS & 13 & 0.71 & 0.92 & 0.71 & 0.92 & 0.71 & 0.92\\
No integrity check & 6 & 0.20 & 0.33 & 1.00 & 0.33 & 1.00 & 1.00\\
\rowcolor{Gray}
Use of weak crypto alg. & 1 & 0.25 & 1.00 & 0.25 & 1.00 & 0.50 & 1.00\\
Missing default case statement & 20 & 1.00 & 0.45 & 1.00 & 0.95 & 1.00 & 0.95\\
\midrule
No smell   & 43 & 0.85 & 0.79 & 0.95 & 0.93 & 0.95 & 0.88\\
\midrule
\textbf{Average}   && 0.49 & 0.60 & 0.77 & 0.76 & 0.77 & 0.86\\
\bottomrule
\end{tabular}
%\caption*{\scriptsize{(N/I - Not implemented, N/A - Not applicable, N/D - No data)}}
\end{table*}

\begin{table*}
\small
\footnotesize
\caption{\GLITCH\ vs SLIC: Accuracy for the Puppet Oracle Datasets \scriptsize{(N/I - Not implemented, N/A - Not applicable, N/D - No data)}}
\label{tab:glitch-slic-oracle-puppet}
\centering
\begin{tabular}[t]{lrrrrrrr}
\toprule
& \multicolumn{7}{c}{\textbf{Original Oracle}} \\
\cmidrule(lr){2-8}
& & \multicolumn{2}{c}{\textbf{SLIC}} & \multicolumn{2}{c}{\textbf{\GLITCH\hspace{0.1cm}(SLIC)}} & \multicolumn{2}{c}{\textbf{\GLITCH}}\\
\cmidrule(lr){3-4} \cmidrule(lr){5-6} \cmidrule(lr){7-8}
\textbf{Smell Name} & \textbf{Occurr.} & \textbf{Precision} & \textbf{Recall} & \textbf{Precision} & \textbf{Recall} & \textbf{Precision} & \textbf{Recall}\\
\midrule
Admin by default & 14 & N/D & 0.00 & N/D & 0.00 & 0.81 & 0.93  \\
\rowcolor{Gray}
Empty password & 5 & 0.60 & 0.60 & 1.00 & 1.00 & 1.00 & 1.00 \\
Hard-coded secret & 11 & 0.10 & 0.73 & 0.14 & 0.82 & 0.14 & 0.82 \\
\rowcolor{Gray}
Invalid IP address binding & 6 & 1.00 & 1.00 & 1.00 & 1.00 & 1.00 & 1.00\\
Suspicious comment & 9 & 0.75 & 1.00 & 0.60 & 1.00 & 0.39 & 1.00\\
\rowcolor{Gray}
Use of HTTP without TLS & 5 & 0.38 & 1.00 & 0.42 & 1.00 & 0.45 & 1.00\\
No integrity check & 1 & N/I & N/I & N/I & N/I & N/D & 0.00\\
\rowcolor{Gray}
Use of weak crypto alg. & 4 & 0.43 & 0.75 & 0.50 & 0.75 & 0.57 & 1.00\\
Missing default case statement & 10 & N/I & N/I & 0.83 & 1.00 & 0.83 & 1.00\\
\midrule
No smell   & 52 & 0.95 & 0.71 & 0.98 & 0.77 & 0.97 & 0.71\\
\midrule
\textbf{Average}   && 0.60 & 0.72 & 0.68 & 0.82 & 0.68 & 0.85\\
\bottomrule
\end{tabular}
%\caption*{\scriptsize{(N/I - Not implemented, N/A - Not applicable, N/D - No data)}}
\end{table*}
%%%%%%%%%%%%%%%%%%%%%%%%%%%%%%%%%%%%%%%%%%%%%%%%%%%%%%%%%%%%%%%%%%%%%%%%%%%%%%%%%%%%%%%%%

\subsection{Security Smells Frequency}
\label{sec:eval:smells-frequency}

%%%%%%%%%%%%%%%%%%%%%%%%%%%%%%%%%%%%%%%%%%%%%%%%%%%%%%%%%%%%%%%%%%%%%%%%%%%%%%%%%%%%%%%%%
% EMPIRICAL STUDY TABLES
%%%%%%%%%%%%%%%%%%%%%%%%%%%%%%%%%%%%%%%%%%%%%%%%%%%%%%%%%%%%%%%%%%%%%%%%%%%%%%%%%%%%%%%%%
\setuldepth{123}

\begin{table*}
\small
\footnotesize
\caption{Smell Occurrences. \scriptsize{(N/I - Not implemented, N/A - Not applicable, N/D - No data)}}
\label{tab:datasets-occurrences}
\centering
\begin{tabular}[t]{lrrrrrrrrrrrr}
\toprule
& & & & & \multicolumn{8}{c}{\textbf{Puppet}}\\
\cmidrule(lr){6-13}
& \multicolumn{2}{c}{\textbf{Ansible}} & \multicolumn{2}{c}{\textbf{Chef}} & \multicolumn{2}{c}{\textbf{GH}} & \multicolumn{2}{c}{\textbf{MOZ}} & \multicolumn{2}{c}{\textbf{OST}} & \multicolumn{2}{c}{\textbf{WIK}}\\
\cmidrule(lr){2-3} \cmidrule(lr){4-5} \cmidrule(lr){6-7} \cmidrule(lr){8-9} \cmidrule(lr){10-11} \cmidrule(lr){12-13}
& \textbf{\scriptsize SLAC} & \textbf{\scriptsize \GLITCH} & \textbf{\scriptsize SLAC} & \textbf{\scriptsize \GLITCH} & \textbf{\scriptsize SLIC} & \textbf{\scriptsize \GLITCH} & \textbf{\scriptsize SLIC} & \textbf{\scriptsize \GLITCH} & \textbf{\scriptsize SLIC} & \textbf{\scriptsize \GLITCH} & \textbf{\scriptsize SLIC} & \textbf{\scriptsize \GLITCH}\\
\midrule
Admin by default & N/I & 10,222 & 248 & 1,821 & 34 & 1,201 & 4 & 30 & 35 & 172 & 6 & 136\\
\rowcolor{Gray}
Empty password & 1,973 & 1,432 & 115 & 303 & 131 & 348 & 20 & 20 & 24 & 127 & 36 & 66\\
Hard-coded secret & 47,735 & 45,325 & 15,100 & 7,763 & 5,608 & 6,236 & 394 & 592 & 1,751 & 2,172 & 858 & 1,114\\
\rowcolor{Gray}
Invalid IP address binding & 914 & 2,033 & 499 & 603 & 179 & 96 & 20 & 26 & 90 & 45 & 41 & 18\\
Suspicious comment & 10,498 & 10,749 & 2,267 & 4,343 & 868 & 1,802 & 202 & 285 & 309 & 965 & 343 & 609\\
\rowcolor{Gray}
Use of HTTP without TLS & 4,812 & 3,393 & 2,507 & 2,281 & 934 & 703 & 52 & 31 & 453 & 163 & 164 & 111\\
No integrity check & 1,146 & 1,359 & 1,662 & 304 & N/I & 44 & N/I & 0 & N/I & 4 & N/I & 3\\
\rowcolor{Gray}
Use of weak crypto alg. & N/I & 1,502 & 76 & 147 & 227 & 109 & 48 & 28 & 27 & 18 & 26 & 21\\
Missing default case statement & N/A & N/A & 702 & 1,890 & N/I & 527 & N/I & 210 & N/I & 36 & N/I & 83\\
\midrule
\textbf{Combined} & 67,078 & 76,015 & 23,176 & 19,455 & 7,981 & 11,066 & 740 & 1,222 & 2,689 & 3,702 & 1,474 & 2,161\\
\bottomrule
\end{tabular}
\end{table*}

\begin{table*}
\small
\footnotesize
\caption{Smell density (per KLOC). \scriptsize{(N/I - Not implemented, N/A - Not applicable, N/D - No data)}}
\label{tab:datasets-density}
\centering
\begin{tabular}[t]{lrrrrrrrrrrrr}
\toprule
& & & & & \multicolumn{8}{c}{\textbf{Puppet}}\\
\cmidrule(lr){6-13}
& \multicolumn{2}{c}{\textbf{Ansible}} & \multicolumn{2}{c}{\textbf{Chef}} & \multicolumn{2}{c}{\textbf{GH}} & \multicolumn{2}{c}{\textbf{MOZ}} & \multicolumn{2}{c}{\textbf{OST}} & \multicolumn{2}{c}{\textbf{WIK}}\\
\cmidrule(lr){2-3} \cmidrule(lr){4-5} \cmidrule(lr){6-7} \cmidrule(lr){8-9} \cmidrule(lr){10-11} \cmidrule(lr){12-13}
& \textbf{\scriptsize SLAC} & \textbf{\scriptsize \GLITCH} & \textbf{\scriptsize SLAC} & \textbf{\scriptsize \GLITCH} & \textbf{\scriptsize SLIC} & \textbf{\scriptsize \GLITCH} & \textbf{\scriptsize SLIC} & \textbf{\scriptsize \GLITCH} & \textbf{\scriptsize SLIC} & \textbf{\scriptsize \GLITCH} & \textbf{\scriptsize SLIC} & \textbf{\scriptsize \GLITCH}\\
\midrule
Admin by default & N/I & 1.97 & 0.04 & 0.30 & 0.06 & 1,97 & 0.06 & 0.45 & 0.16 & 0.79 & 0.04 & 1.00\\
\rowcolor{Gray}
Empty password & 0.38 & 0.28 & 0.02 & 0.05 & 0.21 & 0.57 & 0.30 & 0.30 & 0.11 & 0.58 & 0.27 & 0.49\\
Hard-coded secret & 9.21 & 8.75 & 2.49 & 1.28 & 9.19 & 10.22 & 5.94 & 8.92 & 8.04 & 9.97 & 6.35 & 8.24\\
\rowcolor{Gray}
Invalid IP address binding & 0.18 & 0.39 & 0.08 & 0.10 & 0.29 & 0.16 & 0.30 & 0.39 & 0.41 & 0.21 & 0.30 & 0.13\\
Suspicious comment & 2.03 & 2.07 & 0.37 & 0.72 & 1.42 & 2.95 & 3.04 & 4.29 & 1.42 & 4.43 & 2.54 & 4.51\\
\rowcolor{Gray}
Use of HTTP without TLS & 0.93 & 0.65 & 0.41 & 0.38 & 1.53 & 1.15 & 0.78 & 0.47 & 2.08 & 0.75 & 1.21 & 0.82\\
No integrity check & 0.22 & 0.26 & 0.27 & 0.05 & N/I & 0.07 & N/I & 0.00 & N/I & 0.02 & N/I & 0.02\\
\rowcolor{Gray}
Use of weak crypto alg. & N/I & 0.29 & 0.01 & 0.02 & 0.37 & 0.18 & 0.72 & 0.42 & 0.12 & 0.08 & 0.19 & 0.16\\
Missing default case statement & N/A & N/A & 0.12 & 0.31 & N/I & 0.86 & N/I & 3.16 & N/I & 0.17 & N/I & 0.61\\
\midrule
\textbf{Combined} & 12.95 & 14.66 & 3.81 & 3.21 & 13.07 & 18.13 & 11.14 & 18.40 & 12.34 & 17.00 & 10.90 & 15.98\\
\bottomrule
\end{tabular}
\end{table*}

\begin{table*}
\small
\footnotesize
\caption{Proportion of Scripts (Script\%) with at Least One Smell. \scriptsize{(N/I - Not implemented, N/A - Not applicable, N/D - No data)}}
\label{tab:datasets-proportion}
\centering
\begin{tabular}[t]{lrrrrrrrrrrrr}
\toprule
& & & & & \multicolumn{8}{c}{\textbf{Puppet}}\\
\cmidrule(lr){6-13}
& \multicolumn{2}{c}{\textbf{Ansible}} & \multicolumn{2}{c}{\textbf{Chef}} & \multicolumn{2}{c}{\textbf{GH}} & \multicolumn{2}{c}{\textbf{MOZ}} & \multicolumn{2}{c}{\textbf{OST}} & \multicolumn{2}{c}{\textbf{WIK}}\\
\cmidrule(lr){2-3} \cmidrule(lr){4-5} \cmidrule(lr){6-7} \cmidrule(lr){8-9} \cmidrule(lr){10-11} \cmidrule(lr){12-13}
& \textbf{\scriptsize SLAC} & \textbf{\scriptsize \GLITCH} & \textbf{\scriptsize SLAC} & \textbf{\scriptsize \GLITCH} & \textbf{\scriptsize SLIC} & \textbf{\scriptsize \GLITCH} & \textbf{\scriptsize SLIC} & \textbf{\scriptsize \GLITCH} & \textbf{\scriptsize SLIC} & \textbf{\scriptsize \GLITCH} & \textbf{\scriptsize SLIC} & \textbf{\scriptsize \GLITCH}\\
\midrule
Admin by default & N/I & 5.7 & 0.2 & 1.7 & 0.3 & 3.6 & 0.2 & 1.5 & 1.1 & 5.2 & 0.2 & 4.0\\
\rowcolor{Gray}
Empty password & 0.8 & 0.4 & 0.2 & 0.3 & 1.1 & 2.5 & 0.6 & 0.9 & 0.7 & 3.5 & 0.4 & 1.1\\
Hard-coded secret & 18.3 & 13.9 & 7.7 & 5.2 & 18.2 & 20.2 & 9.9 & 12.3 & 24.6 & 31.3 & 17.0 & 19.1\\
\rowcolor{Gray}
Invalid IP address binding & 0.7 & 0.7 & 0.5 & 0.4 & 1.4 & 0.7 & 0.7 & 0.6 & 2.8 & 1.4 & 1.4 & 0.6\\
Suspicious comment & 5.4 & 5.4 & 2.6 & 3.8 & 5.3 & 8.9 & 8.6 & 11.0 & 7.0 & 13.5 & 9.1 & 13.7\\
\rowcolor{Gray}
Use of HTTP without TLS & 2.3 & 1.6 & 1.8 & 1.8 & 5.1 & 3.7 & 1.5 & 0.9 & 8.2 & 3.1 & 3.8 & 2.5\\
No integrity check & 0.8 & 0.9 & 1.4 & 0.4 & N/I & 0.4 & N/I & 0.0 & N/I & 0.1 & N/I & 0.1\\
\rowcolor{Gray}
Use of weak crypto alg. & N/I & 0.6 & 0.1 & 0.2 & 1.6 & 0.8 & 1.4 & 0.4 & 0.8 & 0.5 & 0.5 & 0.4\\
Missing default case statement & N/A & N/A & 0.9 & 2.1 & N/I & 2.9 & N/I & 9.9 & N/I & 1.0 & N/I & 1.8\\
\midrule
\textbf{Combined} & 23.8 & 19.6 & 11.4 & 10.4 & 25.5 & 29.6 & 18.0 & 27.5 & 32.5 & 40.1 & 26.8 & 31.5 \\
\bottomrule
\end{tabular}
\end{table*}

%%%%%%%%%%%%%%%%%%%%%%%%%%%%%%%%%%%%%%%%%%%%%%%%%%%%%%%%%%%%%%%%%%%%%%%%%%%%%%%%%%%%%%%%%

Using \GLITCH, we performed an empirical study to quantify the prevalence of security smells in Ansible, Chef, and Puppet. Similar studies were performed by Rahman et al.~\cite{rahman2019seven} (for Puppet scripts using SLIC) and Rahman et al.~\cite{rahman2021security} (for Ansible and Chef scripts using SLAC). Here, the goal is to use \GLITCH\ and investigate whether there are any noticeable differences.
The IaC datasets used are described in Section~\ref{sec:eval:datasets} and their attributes shown in Table~\ref{tab:datasets-metrics}.
This means that, when considering the three IaC datasets as a whole, this empirical study considers 1413 repositories with 196,755 IaC scripts. In total, we analyze 12,281,251 LOC.

Similar to previous studies, the first step was to determine the occurrences of security smells for each IaC script. We then calculated the two following metrics:

\begin{itemize}[leftmargin=10pt]
    \item \textbf{Smell density}: frequency of a given security smell for every 1,000 LOC~\cite{kelly1992analysis,rahman2021security}. For a given smell $x$,
    \[
    SmellDensity(x) = \frac{\text{Total occurrences of $x$}}{\text{Total line count for all scripts} / 1000}
    \]
    \item \textbf{Proportion of scripts (Script\%)}: percentage of scripts that contain at least one occurrence of smell $x$.
\end{itemize}
%Tables~\ref{tab:datasets-occurrences},~\ref{tab:datasets-density},~and~\ref{tab:datasets-proportion} present, respectively, the findings on ocurrences, smell density, and proportion of scripts.

\subsubsection{Occurrences}
Looking at Table~\ref{tab:datasets-occurrences}, we observe that all categories of security smells are identified across all datasets.
Overall, \GLITCH\ detects 76,015 security smells for Ansible, 19,455 for Chef, and 18,151 for Puppet. 
%Even though \GLITCH\ supports more types of security smells for Ansible than SLIC (eight vs six), the total number of smells identified is smaller.
\GLITCH\ identifies fewer security smells in the Chef dataset than SLIC.
On the other hand, \GLITCH\ identifies more security smells than SLIC in the Ansible and Puppet datasets  (76,015 vs 67,078 and 18,151 vs 12,884).
When using \GLITCH\ for Ansible and Puppet, the three most dominant security smells are \emph{Hard-coded secret}, \emph{Admin by default}, and \emph{Suspicious comment}. For Chef, the three most dominant security smells are \emph{Hard-coded secret}, \emph{Suspicious comment}, and \emph{Use of HTTP without TLS}.

\subsubsection{Smell density}
Table~\ref{tab:datasets-density} shows the smell density for the three datasets.
Overall, \GLITCH\ detects 14.66 security smells per 1,000 LOC in Ansible scripts, 3.21 in Chef scripts, and an average of 17.38 in Puppet scripts.
For all datasets, the dominant security smell is \emph{Hard-coded secret}, followed by \emph{Suspicious comment}. Given that the precision values for these smells tend to be the lowest (see Section~\ref{sec:eval:accuracy}), this suggests that many of these are false positives. 
% JFF: The following is no longer the case.
%However, there is an exception: for Ansible, the second most dominant smell is \emph{Admin by default}.
The third most dominant security smell differs across the three datasets: for Ansible, it is \emph{Admin by default} (1.97); for Chef, it is \emph{Use of HTTP without TLS} (0.38); and for Puppet, it is \emph{Admin by default} when considering the GitHub dataset (1.97), \emph{Missing default case statement} when considering the Mozilla dataset (3.16), and \emph{Admin by default} when considering the Openstack and Wikimedia datasets (0.79 and 1.00, respectively).

\subsubsection{Proportion of Scripts (Script\%)}
Table~\ref{tab:datasets-proportion} shows, for the three datasets, the proportion of scripts with at least one occurrence of a smell.
For Ansible, \GLITCH\ detects at least one of the eight identified security smells in 19.6\% of the total scripts. For SLIC, the percentage is 23.8\%, but note that SLIC only supports six security smells. This is not very different from the values obtained by Rahman et al.~\cite{rahman2021security}, where the percentages obtained with SLIC were 25.3\% and 29.6\% for their GitHub and Openstack datasets, respectively.
For Chef, \GLITCH\ detects at least one of the nine identified security smells in 10.4\% of the total scripts. For SLIC, the percentage is slightly higher at 11.4\%. Here, we note a more noticeable discrepancy with Rahman et al.'s study~\cite{rahman2021security}: the percentages obtained with SLIC were 20.5\% and 30.4\% for their GitHub and Openstack datasets, respectively.
For Puppet, in the GitHub, Mozilla, OpenStack, and Wikimedia datasets, \GLITCH\ detects at least one of the nine identified security smells in, respectively, 29.6\%, 27.5\%, 40.1\%, and 31.5\% of the total scripts. These percentages are slightly higher than those obtained for SLIC. 
%\remove{Even though we are using the dataset shared by Rahman et al.~\cite{rahman2019seven}, there are very small differences in the percentages obtained in their study: for their GitHub, Mozilla, OpenStack, and Wikimedia datasets, the differences are 0.2\%, 0.1\%, 0.4\%, and 0.1\%.}

For all datasets, the dominant security smell is \emph{Hard-coded secret}, followed by \emph{Suspicious comment}.
Given that the precision values for these smells tend to be the lowest (see Section~\ref{sec:eval:accuracy}), this suggests that many of these are false positives. 
However, there is an exception: for Ansible, the second most dominant smell is \emph{Admin by default} (5.7\%); since the accuracy of \GLITCH\ for this smell is high, this suggests that there is a substantial number of Ansible scripts that are affected by this problem.
The third most dominant security smell differs across the three datasets: for Ansible, it is \emph{Suspicious comment} (5.4\%); for Chef, it is \emph{Missing default case statement} (2.1\%); and for Puppet, it is \emph{Use of HTTP without TLS} when considering the GitHub dataset (3.7\%), \emph{Missing default case statement} when considering the Mozilla dataset (9.9\%), \emph{Admin by default} when considering the Openstack and Wikimedia datasets (5.2\% and 4.0\%, respectively). We note that the high accuracy of \GLITCH\ for the smell \emph{Missing default case statement}, suggests that a substantial number of scripts in the Mozilla dataset are affected by this problem.

\subsubsection{Execution times}
% Intel(R) Core(TM) i5-9400F CPU @ 2.90GHz
% 32 GB RAM
% M.2 PCIE X4 2280 SSD ADATA SPECTRIX S40G RGB 256GB 3500MB/s reading speed 3000MB/s writing speed
% Quantos cores tem isto? 6 cores
% Ubuntu 21.10
% Desktop? Yes
The execution times of \GLITCH, SLIC, and SLAC for the three datasets are shown
in Table~\ref{tab:execution-time} (in seconds). These times were obtained in a server machine running Debian 10, with 4 Intel(R) Xeon(R) CPU E5-2630 v2 @ 2.60GHz, 64GB RAM, and with a Toshiba MG03ACA100 hard drive. We executed 5 runs for each pair tool/dataset and averaged the obtained execution times. Each run was executed in its own Docker container created from the Docker image we provide in the replication package. Runs from the same set of 5 runs were executed simultaneously. \GLITCH\ is much quicker than SLIC and SLAC when running on Chef or Puppet scripts (speedups vary from 9.14$\times$ to 32.07$\times$). SLIC and SLAC respectively call \emph{puppet-lint}\footnote{\url{https://github.com/rodjek/puppet-lint}} and \emph{foodcritic}\footnote{\url{http://www.foodcritic.io/}} to analyze each Puppet or Chef script. The overhead of creating a new system process for each script analyzed and other non-related analyses performed by \emph{puppet-lint} and \emph{foodcritic} are the main reason for the slower execution times. However, when compared to SLAC, \GLITCH\ takes more than double the time to run on the Ansible dataset. This happens because we parse Ansible scripts using \emph{ruamel.yaml}, a Python package slower than the popular \emph{yaml} package, but with the advantage of saving comments in the AST.

\begin{table}
\small
\footnotesize
\caption{The average execution times between 5 runs (seconds).}
\label{tab:execution-time}
\begin{tabular}[t]{lrrrrrrr}
\toprule
& & & \multicolumn{4}{c}{\textbf{Puppet}}\\
\cmidrule(lr){4-7}
\textbf{Tool} & \textbf{Ansible} & \textbf{Chef} & \textbf{GH} & \textbf{MOZ} & \textbf{OST} & \textbf{WIK}\\
\midrule
\textbf{SLIC/SLAC} & 797 & 76,153 & 2,615 & 380 & 915 & 866 \\
\textbf{\GLITCH} & 1,668 & 8,335 & 86 & 14 & 35 & 27\\
\cmidrule(lr){1-7}
\textbf{Speedup} & 0.48 & 9.14 & 30.41 & 27.14 & 26.14 & 32.07\\
\bottomrule
\end{tabular}
\end{table}

%%
%% Discussion
\section{Discussion}
\label{sec:discussion}
In this section, we answer the research questions listed in Section~\ref{sec:eval:rqs}, we discuss the practical implications of our findings, and we outline potential threats to the validity of our work.

\subsection{Answers to Research Questions}
\newcommand{\answer}[2]{{\vspace{0.1cm}\textbf{Answer to RQ#1. }} #2}
Given the findings reported in the previous section, we answer the research questions posed in Section~\ref{sec:eval:rqs} as follows:

\answer{1 [Abstraction]}{\textbf{\sl Can our intermediate representation model
IaC scripts and support automated detection of security smells?}
%\vskip 0.1cm
Yes. We demonstrate that our intermediate representation can model scripts written in different IaC technologies, with our current implementation supporting Ansible, Chef, and Puppet. We also define and implement nine rules that operate on the intermediate representation and that can be used to detect security smells. New rules can be easily created and existing rules can be easily changed. We evaluate our implementation with three large datasets containing 196,755 IaC scripts and 12,281,251 LOC. This strongly suggests that the intermediate representation is robust enough to support a large variety of IaC scripts.}

\answer{2 [Accuracy and Performance]}{\textbf{\sl How does \GLITCH\ compare with existing state-of-art tools for detecting security smells in terms of accuracy and performance?}
%\vskip 0.1cm
As shown in Tables~\ref{tab:glitch-slac-oracle-ansible},~\ref{tab:glitch-slac-oracle-chef},~and~\ref{tab:glitch-slic-oracle-puppet}, the average precision and recall values of \GLITCH\ are substantially better than the average precision and recall values of SLIC and SLAC.
For Puppet, average precision and average recall improved by 8 and 13 percentage points, respectively. For Ansible, average precision improved 10 percentage points and average recall improved 8 percentage points. For Chef, the improvement was more expressive: the average precision and average recall improved by 28 and 26 percentage points, respectively.
In terms of performance, as Table~\ref{tab:execution-time} shows, \GLITCH\ is much faster analyzing Chef and Puppet scripts than tools such as SLIC or SLAC (speedups vary from 9.14$\times$ to 32.07$\times$). For Ansible, \GLITCH\ takes more than twice as long than SLAC, but it can still analyze IaC scripts in an acceptable amount of time (e.g., it took us around 28 minutes to analyze more than 5M LOC).
}

\answer{3 [Frequency]}{\textbf{\sl How frequently do security smells occur in IaC scripts?}
%\vskip 0.1cm
All categories of security smells are identified across all datasets considered in this work.
For Ansible, \GLITCH\ detects at least one of the eight identified security smells in 19.6\% of the total scripts. For Chef, it detects at least one of the nine identified security smells in 10.4\% of the total scripts. For Puppet, in the GitHub, Mozilla, OpenStack, and Wikimedia datasets, \GLITCH\ detects at least one of the nine identified security smells in, respectively, 29.6\%, 27.5\%, 40.1\%, and 31.5\% of the total scripts.

In general, the most dominant security smell is \emph{Hard-coded secret}, followed by \emph{Suspicious comment}. Given that the precision values for these smells tend to be the lowest (see Section~\ref{sec:eval:accuracy}), this suggests that many of these are false positives. 
For Ansible, the second most dominant smell is \emph{Admin by default} (5.7\%). For Chef and for the Mozilla dataset of Puppet scripts, the third most dominant smell is \emph{Missing default case statement} (2.1\% and 9.9\%). Since the accuracy of \GLITCH\ for these smells is high, this suggests that there is a substantial number of Ansible and Chef scripts that are affected by these problems.
}

\subsection{Practical Implications and Challenges}
The main practical implication of this work is that it is now possible to implement new rules to detect code smells that can be immediately applied to a variety of IaC technologies. Some of the rules currently implemented have very high precision and recall, and have been used to identify a considerable number of smells in our study. This suggests that IaC practitioners can benefit if they focus first on smells of those specific categories (e.g., \emph{Admin by default} and \emph{Missing default case statement}). Also, during the development of this work it became clear that there are no open replication packages that IaC researchers and practitioners can use. Therefore, we constructed a new open-source replication package that can be used by the community.

% Challenges
We identify three main challenges: 
%improving the quality of the smell detection rules, extending the scope of the problems and IaC technologies addressed by the current implementation of \GLITCH, and integrating \GLITCH\ into the development process.
%\paragraph{Quality}
\begin{enumerate*}[label=\textbf{(\arabic*)}]
\item {\bfseries Quality. }
This challenge is about increasing the precision and recall of \GLITCH. For example, the definitions of some rules (e.g., those that use many keywords) still report a considerable number of false positives (e.g. \emph{Hard-coded secret}). Future work should be invested in improving the quality of the rules that \GLITCH\ implements. Addressing this challenge is perhaps an important step toward real-life adoption of \GLITCH.

%\paragraph{Scope}
\item {\bfseries Scope. }
This challenge is about extending \GLITCH\ to support more IaC technologies and to detect more vulnerabilities. For example, it would be interesting to extend \GLITCH\ to support Terraform and to support the detection of faults regarding ordering violations~\cite{sotiropoulos2020practical} or intra-update sniping vulnerabilities~\cite{lepiller2021analyzing}. 
%Also, exploring automated repair techniques is an interesting avenue for future work.
By addressing this challenge, we will be in a better position to provide a more precise characterization of the expressiveness of the smell detection engine.
%Comment from reviwewer: it would be good to characterize what type of smells can and cannot be characterized in the intermediate representation; I'd expect there's some loss with the abstraction which makes specifying certain smells unworkable. Are there any limitations in terms of the expressivity of the smell detection engine? For example, it seems to rely heavily on text matching in the "value" fields, how robust/complete would that be?
 
%\paragraph{Development process}
\item {\bfseries Development process. }
This challenge is about integrating these tools into the development process, thus contributing to real-life adoption. %To ease the interaction of developers \GLITCH, thus making it useful during the development life-cycle, the 
The following could bring added value: integration with continuous integration (CI) processes (e.g., GitHub actions), integration with popular IDEs, interactive reports (e.g., highlight vulnerable code), and explainable warnings. Since \GLITCH\ is much faster than other state-of-the-art tools for analyzing Chef and Puppet scripts, it becomes more appealing to integrate \GLITCH\ as part of a CI workflow~\cite{jin2021helped}.
\end{enumerate*}

\subsection{Threats to Validity}
%% Conclusion validity
%% Can we derive convincing theory from the results?
A threat to conclusion validity is that the identification of security smells in the oracle datasets are susceptible to the subjectivity of the raters. We mitigated this by using three raters, with two of them not being authors of the paper and with experience in IaC technologies and/or cybersecurity. Also, we only kept the classifications where at least two raters agreed.

% Construct validity
% Does the experimental model reflect reality?
% TODO? Not include?

%% Internal validity
%% Were the measured improvements caused by the tool?
A threat to internal validity is that, due to the complexity and generality of \GLITCH, there may exist implementation bugs in the codebase. We extensively tested the tool to mitigate this risk. Furthermore, all our code and datasets are publicly available for other researchers and potential users to check the validity of the results. Finally, the detection accuracy of \GLITCH\ depends on the rules that we have provided in Table~\ref{tab:security-rules}. These rules are heuristic-driven and can result in false positives and false negatives.

%% External validity
% How far can we generalize the results?
A threat to external validity is that, since we focus on Ansible, Chef, and Puppet scripts, our findings may not be generalizable to other IaC technologies. Moreover, in its current form, our internal representation might not be rich enough to detect other categories of security smells not considered in this paper. We mitigated this risk by ensuring that the concepts modeled by the intermediate representation are as general as possible and by choosing to demonstrate its validity using three different IaC technologies that, as shown in Table~\ref{tab:iaccharacteristics}, have different characteristics (procedural vs declarative, different configuration setup, etc.). 
Also, the classification of security smells used is subject to practitioner interpretation and their relevance may vary from one practitioner to another. To mitigate this, we followed classifications established by previous work~\cite{rahman2019seven,rahman2021security}.
Finally, all the datasets used in our work are from open-source projects and not from proprietary sources.

% Poderiamos dizer aqui que tb pode nao suportar outras tecnicas, mas talvez possa ficar assim para ja?
% Tirado do projecto de tese: We will implement some techniques found in other studies (e.g. Sotiropoulos et al.'s technique for identifying missing dependencies \cite{sotiropoulos2020practical}

%%
%% Conclusion
\section{Conclusion}
\label{sec:conclusion}
%This paper presents an approach that enables consistent security smell detection across different IaC technologies. We conduct a large-scale empirical study that analyzes security smells on three large datasets containing Ansible, Chef, and Puppet scripts.
%196,756 IaC scripts and 12,281,383 LOC. 
%We show that all categories of security smells are identified across all datasets and we identify some smells that might affect many IaC projects.

This paper shows that it is possible and beneficial to consistently detect security smells across different IaC technologies.
We conducted a large-scale empirical study where we consider nine security smells documented in the literature. We found that all categories of security smells are identified across all datasets. We identified some smells that might affect many IaC projects.
 
An outcome of this work is \GLITCH,
%This paper presents GLITCH, 
a new technology-agnostic framework that allows polyglot security smell detection in IaC scripts, by transforming them into a new intermediate representation on which different security smell detectors can be defined. GLITCH currently supports the detection of nine different security smells 
%and it can analyze scripts written 
in Puppet, Ansible, or Chef scripts. Our evaluation not only shows that GLITCH can reduce the effort of writing security smell analyses for multiple IaC technologies, but also that it has higher precision and recall than the current state-of-the-art tools.

All our code and datasets are publicly available. We argue that \GLITCH\ and the datasets that we created and made available in our replication package are very valuable assets for driving reproducible research in the analysis of IaC scripts.

%%
%% The acknowledgments section is defined using the "acks" environment
%% (and NOT an unnumbered section). This ensures the proper
%% identification of the section in the article metadata, and the
%% consistent spelling of the heading.
\begin{acks}
The authors would like to thank the anonymous reviewers, whose comments and corrections have led to significant improvements.
We would also like to thank Akond Rahman, who very kindly provided access to datasets used in the evaluation of the tools SLIC and SLAC. 
%Also, thanks to Carolina Pereira for her help creating the video demonstrating \GLITCH. 
The first author is funded by the Advanced Computing/EuroCC MSc Fellows Programme, which is funded by EuroHPC under grant agreement No 951732.
This project was supported by national funds through FCT under project UIDB/50021/2020, and by project ANI 045917 funded by
FEDER and FCT.
\end{acks}

%%
%% The next two lines define the bibliography style to be used, and
%% the bibliography file.
\bibliographystyle{ACM-Reference-Format}
\bibliography{references}

%%% -*-BibTeX-*-
%%% Do NOT edit. File created by BibTeX with style
%%% ACM-Reference-Format-Journals [18-Jan-2012].

\begin{thebibliography}{28}

%%% ====================================================================
%%% NOTE TO THE USER: you can override these defaults by providing
%%% customized versions of any of these macros before the \bibliography
%%% command.  Each of them MUST provide its own final punctuation,
%%% except for \shownote{}, \showDOI{}, and \showURL{}.  The latter two
%%% do not use final punctuation, in order to avoid confusing it with
%%% the Web address.
%%%
%%% To suppress output of a particular field, define its macro to expand
%%% to an empty string, or better, \unskip, like this:
%%%
%%% \newcommand{\showDOI}[1]{\unskip}   % LaTeX syntax
%%%
%%% \def \showDOI #1{\unskip}           % plain TeX syntax
%%%
%%% ====================================================================

\ifx \showCODEN    \undefined \def \showCODEN     #1{\unskip}     \fi
\ifx \showDOI      \undefined \def \showDOI       #1{#1}\fi
\ifx \showISBNx    \undefined \def \showISBNx     #1{\unskip}     \fi
\ifx \showISBNxiii \undefined \def \showISBNxiii  #1{\unskip}     \fi
\ifx \showISSN     \undefined \def \showISSN      #1{\unskip}     \fi
\ifx \showLCCN     \undefined \def \showLCCN      #1{\unskip}     \fi
\ifx \shownote     \undefined \def \shownote      #1{#1}          \fi
\ifx \showarticletitle \undefined \def \showarticletitle #1{#1}   \fi
\ifx \showURL      \undefined \def \showURL       {\relax}        \fi
% The following commands are used for tagged output and should be
% invisible to TeX
\providecommand\bibfield[2]{#2}
\providecommand\bibinfo[2]{#2}
\providecommand\natexlab[1]{#1}
\providecommand\showeprint[2][]{arXiv:#2}

\bibitem[Alnafessah et~al\mbox{.}(2021)]%
        {alnafessah2021quality}
\bibfield{author}{\bibinfo{person}{Ahmad Alnafessah}, \bibinfo{person}{Alim~Ul
  Gias}, \bibinfo{person}{Runan Wang}, \bibinfo{person}{Lulai Zhu},
  \bibinfo{person}{Giuliano Casale}, {and} \bibinfo{person}{Antonio Filieri}.}
  \bibinfo{year}{2021}\natexlab{}.
\newblock \showarticletitle{Quality-Aware DevOps Research: Where Do We Stand?}
\newblock \bibinfo{journal}{\emph{IEEE Access}}  \bibinfo{volume}{9}
  (\bibinfo{year}{2021}), \bibinfo{pages}{44476--44489}.
\newblock


\bibitem[Fryman(2014)]%
        {fryman_2014}
\bibfield{author}{\bibinfo{person}{James Fryman}.}
  \bibinfo{year}{2014}\natexlab{}.
\newblock \bibinfo{title}{DNS outage post mortem}.
\newblock
\newblock
\urldef\tempurl%
\url{https://github.blog/2014-01-18-dns-outage-post-mortem/}
\showURL{%
\tempurl}
\newblock
\shownote{Accessed: 3 May 2022}.


\bibitem[Guerriero et~al\mbox{.}(2019)]%
        {guerriero2019adoption}
\bibfield{author}{\bibinfo{person}{Michele Guerriero}, \bibinfo{person}{Martin
  Garriga}, \bibinfo{person}{Damian~A Tamburri}, {and} \bibinfo{person}{Fabio
  Palomba}.} \bibinfo{year}{2019}\natexlab{}.
\newblock \showarticletitle{Adoption, support, and challenges of
  infrastructure-as-code: Insights from industry}. In
  \bibinfo{booktitle}{\emph{2019 IEEE International Conference on Software
  Maintenance and Evolution (ICSME)}}. IEEE, \bibinfo{pages}{580--589}.
\newblock


\bibitem[Hanappi et~al\mbox{.}(2016)]%
        {hanappi2016asserting}
\bibfield{author}{\bibinfo{person}{Oliver Hanappi}, \bibinfo{person}{Waldemar
  Hummer}, {and} \bibinfo{person}{Schahram Dustdar}.}
  \bibinfo{year}{2016}\natexlab{}.
\newblock \showarticletitle{Asserting reliable convergence for configuration
  management scripts}. In \bibinfo{booktitle}{\emph{Proceedings of the 2016 ACM
  SIGPLAN International Conference on Object-Oriented Programming, Systems,
  Languages, and Applications}}. \bibinfo{pages}{328--343}.
\newblock


\bibitem[Hersher(2017)]%
        {hersher_2017}
\bibfield{author}{\bibinfo{person}{Rebecca Hersher}.}
  \bibinfo{year}{2017}\natexlab{}.
\newblock \bibinfo{title}{Amazon and the \$150 Million typo}.
\newblock
\newblock
\urldef\tempurl%
\url{https://www.npr.org/sections/thetwo-way/2017/03/03/518322734/amazon-and-the-150-million-typo?t=1651588365675}
\showURL{%
\tempurl}
\newblock
\shownote{Accessed: 3 May 2022}.


\bibitem[Ikeshita et~al\mbox{.}(2017)]%
        {ikeshita2017test}
\bibfield{author}{\bibinfo{person}{Katsuhiko Ikeshita}, \bibinfo{person}{Fuyuki
  Ishikawa}, {and} \bibinfo{person}{Shinichi Honiden}.}
  \bibinfo{year}{2017}\natexlab{}.
\newblock \showarticletitle{Test suite reduction in idempotence testing of
  infrastructure as code}. In \bibinfo{booktitle}{\emph{International
  Conference on Tests and Proofs}}. Springer, \bibinfo{pages}{98--115}.
\newblock


\bibitem[Jiang and Adams(2015)]%
        {jiang2015co}
\bibfield{author}{\bibinfo{person}{Yujuan Jiang} {and} \bibinfo{person}{Bram
  Adams}.} \bibinfo{year}{2015}\natexlab{}.
\newblock \showarticletitle{Co-evolution of infrastructure and source code-an
  empirical study}. In \bibinfo{booktitle}{\emph{2015 IEEE/ACM 12th Working
  Conference on Mining Software Repositories}}. IEEE, \bibinfo{pages}{45--55}.
\newblock


\bibitem[Jin and Servant(2021)]%
        {jin2021helped}
\bibfield{author}{\bibinfo{person}{Xianhao Jin} {and}
  \bibinfo{person}{Francisco Servant}.} \bibinfo{year}{2021}\natexlab{}.
\newblock \showarticletitle{What helped, and what did not? An Evaluation of the
  Strategies to Improve Continuous Integration}. In
  \bibinfo{booktitle}{\emph{2021 IEEE/ACM 43rd International Conference on
  Software Engineering (ICSE)}}. IEEE, \bibinfo{pages}{213--225}.
\newblock


\bibitem[Kelly et~al\mbox{.}(1992)]%
        {kelly1992analysis}
\bibfield{author}{\bibinfo{person}{John~C Kelly}, \bibinfo{person}{Joseph~S
  Sherif}, {and} \bibinfo{person}{Jonathan Hops}.}
  \bibinfo{year}{1992}\natexlab{}.
\newblock \showarticletitle{An analysis of defect densities found during
  software inspections}.
\newblock \bibinfo{journal}{\emph{Journal of Systems and Software}}
  \bibinfo{volume}{17}, \bibinfo{number}{2} (\bibinfo{year}{1992}),
  \bibinfo{pages}{111--117}.
\newblock


\bibitem[Lepiller et~al\mbox{.}(2021)]%
        {lepiller2021analyzing}
\bibfield{author}{\bibinfo{person}{Julien Lepiller}, \bibinfo{person}{Ruzica
  Piskac}, \bibinfo{person}{Martin Sch{\"a}f}, {and} \bibinfo{person}{Mark
  Santolucito}.} \bibinfo{year}{2021}\natexlab{}.
\newblock \showarticletitle{Analyzing Infrastructure as Code to Prevent
  Intra-update Sniping Vulnerabilities.}. In \bibinfo{booktitle}{\emph{TACAS
  (2)}}. \bibinfo{pages}{105--123}.
\newblock


\bibitem[{MITRE}(2022)]%
        {mitre2022}
\bibfield{author}{\bibinfo{person}{{MITRE}}.} \bibinfo{year}{2022}\natexlab{}.
\newblock \bibinfo{title}{{CWE-Common Weakness Enumeration}}.
\newblock
\newblock
\newblock
\shownote{\url{https://cwe.mitre.org/index.html}}.


\bibitem[Munaiah et~al\mbox{.}(2017)]%
        {munaiah2017curating}
\bibfield{author}{\bibinfo{person}{Nuthan Munaiah}, \bibinfo{person}{Steven
  Kroh}, \bibinfo{person}{Craig Cabrey}, {and} \bibinfo{person}{Meiyappan
  Nagappan}.} \bibinfo{year}{2017}\natexlab{}.
\newblock \showarticletitle{Curating github for engineered software projects}.
\newblock \bibinfo{journal}{\emph{Empirical Software Engineering}}
  \bibinfo{volume}{22}, \bibinfo{number}{6} (\bibinfo{year}{2017}),
  \bibinfo{pages}{3219--3253}.
\newblock


\bibitem[Mutaf(1999)]%
        {mutaf1999defending}
\bibfield{author}{\bibinfo{person}{Pars Mutaf}.}
  \bibinfo{year}{1999}\natexlab{}.
\newblock \showarticletitle{Defending against a Denial-of-Service Attack on
  TCP.}. In \bibinfo{booktitle}{\emph{Recent Advances in Intrusion Detection}}.
\newblock


\bibitem[{National Institute of Standards and Technology}(2014)]%
        {nist2014}
\bibfield{author}{\bibinfo{person}{{National Institute of Standards and
  Technology}}.} \bibinfo{year}{2014}\natexlab{}.
\newblock \bibinfo{title}{{Security and Privacy Controls for Federal
  Information Systems and Organizations}}.
\newblock
\newblock
\newblock
\shownote{\url{https://www.nist.gov/publications/security-and-privacy-controls-federal-information-systems-and-organizations-including-0}}.


\bibitem[Rahman et~al\mbox{.}(2020b)]%
        {rahman2020gang}
\bibfield{author}{\bibinfo{person}{Akond Rahman}, \bibinfo{person}{Effat
  Farhana}, \bibinfo{person}{Chris Parnin}, {and} \bibinfo{person}{Laurie
  Williams}.} \bibinfo{year}{2020}\natexlab{b}.
\newblock \showarticletitle{Gang of eight: A defect taxonomy for infrastructure
  as code scripts}. In \bibinfo{booktitle}{\emph{2020 IEEE/ACM 42nd
  International Conference on Software Engineering (ICSE)}}. IEEE,
  \bibinfo{pages}{752--764}.
\newblock


\bibitem[Rahman et~al\mbox{.}(2020a)]%
        {rahman2020code}
\bibfield{author}{\bibinfo{person}{Akond Rahman}, \bibinfo{person}{Effat
  Farhana}, {and} \bibinfo{person}{Laurie Williams}.}
  \bibinfo{year}{2020}\natexlab{a}.
\newblock \showarticletitle{The ‘as code’activities: development
  anti-patterns for infrastructure as code}.
\newblock \bibinfo{journal}{\emph{Empirical Software Engineering}}
  \bibinfo{volume}{25}, \bibinfo{number}{5} (\bibinfo{year}{2020}),
  \bibinfo{pages}{3430--3467}.
\newblock


\bibitem[Rahman et~al\mbox{.}(2019a)]%
        {rahman2019systematic}
\bibfield{author}{\bibinfo{person}{Akond Rahman}, \bibinfo{person}{Rezvan
  Mahdavi-Hezaveh}, {and} \bibinfo{person}{Laurie Williams}.}
  \bibinfo{year}{2019}\natexlab{a}.
\newblock \showarticletitle{A systematic mapping study of infrastructure as
  code research}.
\newblock \bibinfo{journal}{\emph{Information and Software Technology}}
  \bibinfo{volume}{108} (\bibinfo{year}{2019}), \bibinfo{pages}{65--77}.
\newblock


\bibitem[Rahman et~al\mbox{.}(2019b)]%
        {rahman2019seven}
\bibfield{author}{\bibinfo{person}{Akond Rahman}, \bibinfo{person}{Chris
  Parnin}, {and} \bibinfo{person}{Laurie Williams}.}
  \bibinfo{year}{2019}\natexlab{b}.
\newblock \showarticletitle{The seven sins: Security smells in infrastructure
  as code scripts}. In \bibinfo{booktitle}{\emph{2019 IEEE/ACM 41st
  International Conference on Software Engineering (ICSE)}}. IEEE,
  \bibinfo{pages}{164--175}.
\newblock


\bibitem[Rahman et~al\mbox{.}(2021)]%
        {rahman2021security}
\bibfield{author}{\bibinfo{person}{Akond Rahman}, \bibinfo{person}{Md~Rayhanur
  Rahman}, \bibinfo{person}{Chris Parnin}, {and} \bibinfo{person}{Laurie
  Williams}.} \bibinfo{year}{2021}\natexlab{}.
\newblock \showarticletitle{Security smells in ansible and chef scripts: A
  replication study}.
\newblock \bibinfo{journal}{\emph{ACM Transactions on Software Engineering and
  Methodology (TOSEM)}} \bibinfo{volume}{30}, \bibinfo{number}{1}
  (\bibinfo{year}{2021}), \bibinfo{pages}{1--31}.
\newblock


\bibitem[Rahman and Williams(2018)]%
        {rahman2018characterizing}
\bibfield{author}{\bibinfo{person}{Akond Rahman} {and} \bibinfo{person}{Laurie
  Williams}.} \bibinfo{year}{2018}\natexlab{}.
\newblock \showarticletitle{Characterizing defective configuration scripts used
  for continuous deployment}. In \bibinfo{booktitle}{\emph{2018 IEEE 11th
  International conference on software testing, verification and validation
  (ICST)}}. IEEE, \bibinfo{pages}{34--45}.
\newblock


\bibitem[Rahman and Williams(2019)]%
        {rahman2019source}
\bibfield{author}{\bibinfo{person}{Akond Rahman} {and} \bibinfo{person}{Laurie
  Williams}.} \bibinfo{year}{2019}\natexlab{}.
\newblock \showarticletitle{Source code properties of defective infrastructure
  as code scripts}.
\newblock \bibinfo{journal}{\emph{Information and Software Technology}}
  \bibinfo{volume}{112} (\bibinfo{year}{2019}), \bibinfo{pages}{148--163}.
\newblock


\bibitem[Rescorla et~al\mbox{.}(2000)]%
        {rescorla2000http}
\bibfield{author}{\bibinfo{person}{Eric Rescorla} {et~al\mbox{.}}}
  \bibinfo{year}{2000}\natexlab{}.
\newblock \bibinfo{title}{HTTP over TLS}.
\newblock
\newblock
\newblock
\shownote{RFC 2818, May}.


\bibitem[Salda{\~n}a(2021)]%
        {saldana2021coding}
\bibfield{author}{\bibinfo{person}{Johnny Salda{\~n}a}.}
  \bibinfo{year}{2021}\natexlab{}.
\newblock \bibinfo{booktitle}{\emph{The coding manual for qualitative
  researchers}}.
\newblock \bibinfo{publisher}{sage}.
\newblock


\bibitem[Schwarz et~al\mbox{.}(2018)]%
        {schwarz2018code}
\bibfield{author}{\bibinfo{person}{Julian Schwarz}, \bibinfo{person}{Andreas
  Steffens}, {and} \bibinfo{person}{Horst Lichter}.}
  \bibinfo{year}{2018}\natexlab{}.
\newblock \showarticletitle{Code smells in infrastructure as code}. In
  \bibinfo{booktitle}{\emph{2018 11th International Conference on the Quality
  of Information and Communications Technology (QUATIC)}}. IEEE,
  \bibinfo{pages}{220--228}.
\newblock


\bibitem[Shambaugh et~al\mbox{.}(2016)]%
        {shambaugh2016rehearsal}
\bibfield{author}{\bibinfo{person}{Rian Shambaugh}, \bibinfo{person}{Aaron
  Weiss}, {and} \bibinfo{person}{Arjun Guha}.} \bibinfo{year}{2016}\natexlab{}.
\newblock \showarticletitle{Rehearsal: A configuration verification tool for
  puppet}. In \bibinfo{booktitle}{\emph{Proceedings of the 37th ACM SIGPLAN
  Conference on Programming Language Design and Implementation}}.
  \bibinfo{pages}{416--430}.
\newblock


\bibitem[Sharma et~al\mbox{.}(2016)]%
        {sharma2016does}
\bibfield{author}{\bibinfo{person}{Tushar Sharma}, \bibinfo{person}{Marios
  Fragkoulis}, {and} \bibinfo{person}{Diomidis Spinellis}.}
  \bibinfo{year}{2016}\natexlab{}.
\newblock \showarticletitle{Does your configuration code smell?}. In
  \bibinfo{booktitle}{\emph{2016 IEEE/ACM 13th Working Conference on Mining
  Software Repositories (MSR)}}. IEEE, \bibinfo{pages}{189--200}.
\newblock


\bibitem[Sotiropoulos et~al\mbox{.}(2020)]%
        {sotiropoulos2020practical}
\bibfield{author}{\bibinfo{person}{Thodoris Sotiropoulos},
  \bibinfo{person}{Dimitris Mitropoulos}, {and} \bibinfo{person}{Diomidis
  Spinellis}.} \bibinfo{year}{2020}\natexlab{}.
\newblock \showarticletitle{{Practical fault detection in Puppet programs}}. In
  \bibinfo{booktitle}{\emph{Proceedings of the ACM/IEEE 42nd International
  Conference on Software Engineering}}. \bibinfo{pages}{26--37}.
\newblock


\bibitem[Van~der Bent et~al\mbox{.}(2018)]%
        {van2018good}
\bibfield{author}{\bibinfo{person}{Eduard Van~der Bent},
  \bibinfo{person}{Jurriaan Hage}, \bibinfo{person}{Joost Visser}, {and}
  \bibinfo{person}{Georgios Gousios}.} \bibinfo{year}{2018}\natexlab{}.
\newblock \showarticletitle{How good is your puppet? an empirically defined and
  validated quality model for puppet}. In \bibinfo{booktitle}{\emph{2018 IEEE
  25th international conference on software analysis, evolution and
  reengineering (SANER)}}. IEEE, \bibinfo{pages}{164--174}.
\newblock


\end{thebibliography}

\end{document}